\documentclass[aps,pre,superscriptaddress,floatfix]{revtex4}
\usepackage{graphicx}
\usepackage{pstcol,pst-plot}

\renewcommand\bottomfraction\topfraction
\makeatletter
\newcommand\tr{\mathop{\operator@font tr}}
\makeatother

\begin{document}
\title{Quantization of Classical Maps with tunable Ruelle-Pollicott
  Resonances}
\author{Andrzej Ostruszka}
  \affiliation{Instytut Fizyki, Uniwersytet Jagiello\'nski,\\
    ul.~Reymonta 4,  30-059 Krak\'ow, Poland}
\author{Christopher Manderfeld}
  \affiliation{Fachbereich Physik, Universit\"at Duisburg-Essen, 45117
    Essen, Germany}
\author{Karol \.Zyczkowski}
  \affiliation{Instytut Fizyki, Uniwersytet Jagiello\'nski,\\
    ul.~Reymonta 4,  30-059 Krak\'ow, Poland}
  \affiliation{Centrum Fizyki Teoretycznej, Polska Akademia Nauk,\\
    al.~Lotnik\'ow 32, 02-668 Warszawa, Poland}
\author{Fritz Haake}
  \affiliation{Fachbereich Physik, Universit\"at Duisburg-Essen, 45117
    Essen, Germany}
\date{May 7, 2003}
\begin{abstract}
  We investigate the correspondence between the decay of correlation in
  classical systems, governed by Ruelle--Pollicott resonances, and the
  properties of the corresponding quantum systems.  For this purpose we
  construct classical dynamics with controllable resonances together with
  their quantum counterparts. As an application of such tailormade
  resonances we reveal the role of Ruelle--Pollicott resonances for the
  localization properties of quantum energy eigenstates. 
\end{abstract}

\maketitle

\newcommand{\jinft}{\mbox{$
\,{}_{{}_{N\rightarrow\infty}}
\!\!\!\!\!\!\!\!\!\!\!\longrightarrow\,$}}
\newcommand{\lb}{\mbox{\bf (}}
\newcommand{\rb}{\mbox{\bf )}}
\newcommand{\mb}{\mbox{$|\!\!|$}}

\section{Introduction}

The quantum-classical correspondence of non-integrable systems has been
studied for a long time.  In recent years the role of classical
Ruelle--Pollicott resonances for the dynamics of the quantum counterparts
has become a~point of interest \cite{Altshuler,Zirnbauer,Pance,chris1}.
The classical time evolution can be described in the Liouville picture
as the propagation of the phase-space density $\rho(x,t)$, 
where $x$ denotes a point in phase space.  The
corresponding propagator $P$ is called Frobenius--Perron (FP) operator
and can be defined by
\begin{equation}\label{eq:FPdef}
  \rho(x,t) = \int {\rm d}x'\,\delta(x-F^t(x'))\rho(x',0) \equiv P^t\rho(x,0)\ ,
\end{equation}
where $F$ is the flow in the phase space generated by the dynamics
$x(t) = F^t(x(0))$.  The poles of the resolvent of this operator are the
Ruelle--Pollicott resonances~\cite{Ruelle,Pollicott}.  It turns out that
these resonances correspond to decay rates of classical correlation
functions describing the relaxation process in a chaotic
system~\cite{GaspardBook}.  The presence of the Ruelle--Pollicott
resonances related to slow decay of correlations can
explain non-universal behaviour of the corresponding quantum system,
i.e.~deviations from predictions of the random-matrix theory (RMT)
analyzed e.g. in~\cite{HaakeBook}.

In order to reveal such effects of resonances 
we first construct a classical system with
an isolated, controllable resonance which can be computed
analytically.  We focus our considerations on dynamical systems with
compact phase spaces, in particular the unit sphere.  Periodic
driving destroys integrability, where the stroboscopic description of such
a dynamics is given by a Hamiltonian map.  For this case the
Ruelle--Pollicott resonances are located inside the unit circle of the
complex plane, while decay rates are related to the moduli of
resonances.  For the quantum counterpart the stroboscopic description of
the propagation of wave functions is given by a~unitary Floquet
operator.  The eigenphases of that operator are also called
quasi-eigenenergies. 

Analytical calculations of Ruelle-Pollicott resonances are feasible 
for purely hyperbolic systems~\cite{ChaosBook,Nonnenm}. 
We introduce dynamics which are not Hamiltonian (continuous) but
still area preserving, coupled baker maps on the sphere.  
These model systems are
introduced in Sec.~\ref{sec:class}, where we also show how to find
their periodic orbits, approximate resonances and calculate the traces
of the Frobenius--Perron operators associated with them.  Construction
of the corresponding quantum propagators together with the comparison of
quantum and classical dynamics is presented in Sec.~\ref{sec:quant}.  In
Sec.~\ref{sec:overlaps} we investigate how Ruelle--Pollicott resonances
give rise to the deviations from the random--matrix theory.  Eventually
in Sec.~\ref{sec:crm} we present the model of coupled random matrices which
can be considered as a~simplification of the systems introduced in
Sec.~\ref{sec:class}.

\section{Coupled baker maps}\label{sec:class}

We are interested in classical dynamical systems for which there exists
a~Ruelle--Pollicott resonance with large modulus.  Such
a~resonance governs the long-time behaviour of the system and 
is easy to detect.

The idea standing behind our model systems is rather simple.  Suppose
our system is initially composed out of two disconnected subsystems.  
An arbitrary initial density placed in one of those subsystems
will not spread into
the other subsystem.  The system as a~whole will thus have two invariant
(stationary) densities, 
one for each subsystem (and the linear combinations thereof).
This fact will be reflected in the spectrum of the Frobenius--Perron
operator as a~doubly degenerated eigenvalue $\lambda_1=\lambda_2=1$.
However, if we introduce a~small coupling between both subsystems, the
density from one subsystem will slowly leak into the other one and
eventually reach the invariant density of the entire system.  As
a~result of the coupling the degeneracy of the spectrum will be lifted.
The largest eigenvalue $\lambda_1=1$ corresponds to the unique invariant
density of the entire system, while the other eigenvalue $\lambda_2$
with $|\lambda_2|=1-\epsilon<1$ corresponds to a~metastable state.  The
smaller the spectral gap $\epsilon$, the longer the decay time of the state.

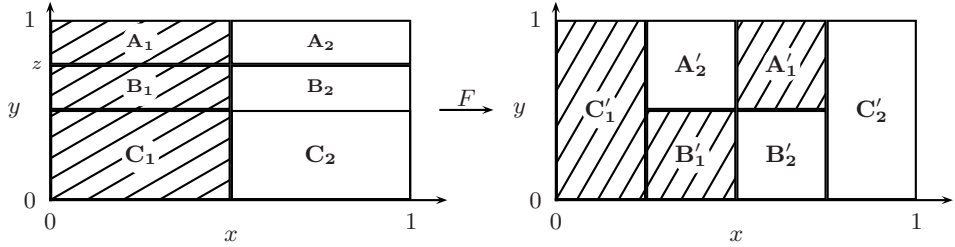
\begin{figure}
  \centering
  \begin{pspicture}(-.5,-.5)(12.5,2.5)
  \psset{yunit=2.4\psunit,xunit=4.8\psunit,hatchsep=6pt}
  \rput[lb](0,0){
    \psaxes[ticks=none]{->}(0,0)(1.1,1.1)
      \rput(-.1,.5){$y$} \rput(.5,-.2){$x$}
      \uput{3pt}[l](0,.75){\scriptsize$z$}
    \psset{fillstyle=none}
    \psframe(.5,0)(1,.5) \psframe(.5,0)(1,.75) \psframe(.5,.75)(1,1)
      \rput(.75,.25){$\mathbf{C_2}$} \rput(.75,.625){\scriptsize$\mathbf{B_2}$}
      \rput(.75,.875){\scriptsize$\mathbf{A_2}$}
    \psset{hatchangle=30,fillstyle=hlines}
    \psframe(0,0)(.5,.5) \psframe(0,.5)(.5,.75) \psframe(0,.75)(.5,1)
      \psset{linecolor=white}
      \qdisk(.25, .25){0.25}\rput(.25,.25){$\mathbf{C_1}$}
      \qdisk(.25,.625){0.25}\rput(.25,.625){\scriptsize$\mathbf{B_1}$}
      \qdisk(.25,.875){0.25}\rput(.25,.875){\scriptsize$\mathbf{A_1}$}
  }
  \psline{->}(1.1,.5)(1.25,.5) \rput[b](1.175,.52){$F$}
  \rput[lb](1.4,0){
    \psaxes[ticks=none]{->}(0,0)(1.1,1.1)
      \rput(-.1,.5){$y$} \rput(.5,-.2){$x$}
    \psset{fillstyle=none}
    \psframe(.75,0)(1,1) \psframe(.5,0)(.75,.5) \psframe(.25,.5)(.5,1)
      \rput(.875,.5){$\mathbf{C'_2}$} \rput(.625,.25){$\mathbf{B'_2}$}
      \rput(.375,.75){$\mathbf{A'_2}$}
    \psset{hatchangle=60,fillstyle=hlines}
    \psframe(0,0)(.25,1) \psframe(.25,0)(.5,.5) \psframe(.5,.5)(.75,1)
      \psset{linecolor=white}
      \qdisk(.125,.5){0.25}\rput(.125,.5){$\mathbf{C'_1}$}
      \qdisk(.375,.25){0.25}\rput(.375,.25){$\mathbf{B'_1}$}
      \qdisk(.625,.75){0.25}\rput(.625,.75){$\mathbf{A'_1}$}
  }
  \end{pspicture}
  \caption{Map on the unit square defined by~(\ref{eq:BNdef}) for $z=3/4$.}
  \label{fig:BNdef}
\end{figure}

For the internal dynamics of both subsystems we choose the standard
baker map acting on a~unit square --- a~well known model of chaotic
dynamics~\cite{Arnold}.  One possible way to introduce the coupling is
described by
\begin{equation}\label{eq:BNdef}
  (y',x') = F(y,x) = \left\{ \begin{array}{ll}
    (2y-1,x/2+1/2)      & y\geq z,x\leq1/2 \\
    (2y-1,x/2+1/4)      & y\in[1/2,z),x\leq1/2 \\
    (2y,x/2)            & y<1/2,x\leq1/2 \\
    (2y-1,x/2)          & y\geq z,x>1/2 \\
    (2y-1,x/2+1/4)      & y\in[1/2,z),x>1/2 \\
    (2y,x/2+1/2)        & y<1/2,x>1/2 \\
    \end{array}\right.\ ,
\end{equation}
whose action is depicted in Fig.~\ref{fig:BNdef}.
In the limiting case $z=1$ the parts $A_1$ and $A_2$ vanish and there
exist two separated subsystems, with indices 1~and~2 (each of them
equivalent to the standard baker map).  For $z<1$ both subsystems are
coupled together.  To describe the strength of the coupling we introduce
a~parameter $\Delta=1-z$ which varies from $0$ to $1/2$. 

\subsection{The system with a~negative coupling}
\label{sec:neg_coupl}

We now present a~slightly modified version of the map~(\ref{eq:BNdef})
which results in the resonance of a~large modulus with negative
sign.  Thus we call both versions of the model as ``positive'' and
``negative'' coupling depending on the sign of the resonance.

We start again with two uncoupled baker maps.  In addition to their
internal dynamics, we assume that in every iteration of the map both
subsystems exchange their positions.  The FP operator corresponding to
such a~system will thus have two eigenvalues of unit modulus:
$\lambda_1=+1$ and $\lambda_2=-1$.  Any
small coupling of both subsystems will cause the density
placed in one subsystem to slowly leak into the other one so the entire
system will possess a~resonance of large modulus and its negative sign
will reflect oscillatory nature of the system.

\begin{figure}
  \centering
  \begin{pspicture}(-.5,-.5)(12.5,2.5)
  \psset{yunit=2.4\psunit,xunit=4.8\psunit,hatchsep=6pt}
  \rput[lb](0,0){
    \psaxes[ticks=none]{->}(0,0)(1.1,1.1)
      \rput(-.1,.5){$y$} \rput(.5,-.2){$x$}
      \uput{3pt}[l](0,.25){\scriptsize$z$}
    \psset{fillstyle=none}
    \psframe(.5,.5)(1,1) \psframe(.5,.25)(1,.5) \psframe(.5,0)(1,.25)
      \rput(.75,.75){$\mathbf{A_2}$}
      \rput(.75,.375){\scriptsize$\mathbf{B_2}$}
      \rput(.75,.125){\scriptsize$\mathbf{C_2}$}
    \psset{hatchangle=30,fillstyle=hlines}
    \psframe(0,.5)(.5,1) \psframe(0,.25)(.5,.5) \psframe(0,0)(.5,.25)
      \psset{linecolor=white}
      \qdisk(.25,.75){0.25}\rput(.25,.75){$\mathbf{A_1}$}
      \qdisk(.25,.375){0.25}\rput(.25,.375){\scriptsize$\mathbf{B_1}$}
      \qdisk(.25,.125){0.25}\rput(.25,.125){\scriptsize$\mathbf{C_1}$}
  }
  \psline{->}(1.1,.5)(1.25,.5) \rput[b](1.175,.52){$F$}
  \rput[lb](1.4,0){
    \psaxes[ticks=none]{->}(0,0)(1.1,1.1)
      \rput(-.1,.5){$y$} \rput(.5,-.2){$x$}
    \psset{fillstyle=none}
    \psframe(.75,0)(1,.5) \psframe(0,.5)(.25,1) \psframe(.25,0)(.5,1)
      \rput(.875,.25){$\mathbf{C'_2}$} \rput(.125,.75){$\mathbf{B'_2}$}
      \rput(.375,.5){$\mathbf{A'_2}$}
    \psset{hatchangle=60,fillstyle=hlines}
    \psframe(0,0)(.25,.5) \psframe(.75,.5)(1,1) \psframe(.5,0)(.75,1)
      \psset{linecolor=white}
      \qdisk(.125,.25){0.25}\rput(.125,.25){$\mathbf{C'_1}$}
      \qdisk(.875,.75){0.25}\rput(.875,.75){$\mathbf{B'_1}$}
      \qdisk(.625,.5){0.25}\rput(.625,.5){$\mathbf{A'_1}$}
  }
  \end{pspicture}
  \caption{Action of the map~(\ref{eq:BSdef}) for $z=1/4$.}
  \label{fig:BSdef}
\end{figure}
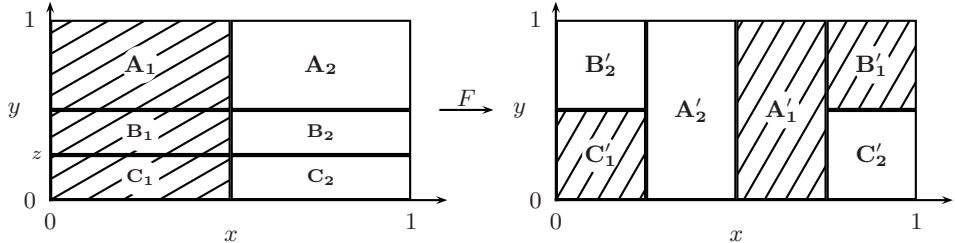

The version of the coupling that we have chosen is presented in
Fig.~\ref{fig:BSdef} and is defined by
\begin{equation}\label{eq:BSdef}
  (y',x') = F(y,x) = \left\{ \begin{array}{ll}
    (2y-1,x/2+1/2)      & y\geq 1/2,x\leq1/2 \\
    (2y,x/2+3/4)        & y\in[z,1/2),x\leq1/2 \\
    (2y,x/2)            & y<z,x\leq1/2 \\
    (2y-1,x/2)          & y\geq 1/2,x>1/2 \\
    (2y,x/2-1/4)        & y\in[z,1/2),x>1/2 \\
    (2y,x/2+1/2)        & y<z,x>1/2 \\
    \end{array}\right.\ .
\end{equation}
Accordingly for this system the coupling strength parameter is given by
$\Delta=z$.  The full coupling ($\Delta=1/2$) limits of systems
(\ref{eq:BNdef}) and (\ref{eq:BSdef}) coincide and correspond to the
baker map on the sphere defined in~\cite{Pakonski1999}.

Our ultimate goal is to construct a~quantum system corresponding to the
classical system with a~large resonance and to investigate the
influence of the classical resonance on the properties of the quantum
system.  For this purpose we
make use of periodic orbits of the classical system and the spectrum of
the Frobenius--Perron operator corresponding to the classical system ---
which we will approximate by introduction of a~stochastic perturbation
into the system.  Hence
the following two sections are devoted to these subjects.

\subsection{Periodic orbits of the classical system}\label{sec:fixpoints}

For the determination of the periodic orbits of the system we will use
the Markov partition.  For any dynamical system $F$ it is defined as
a~such partition $\mathcal{C}$ of the
phase space into $K$ cells 
that the borders of each cell are composed of segments of stable and
unstable manifolds of the system.  Additionally this partition has to
fulfill
\begin{eqnarray}
  F(\partial_s\mathcal{C}) & \subset & \partial_s\mathcal{C},\\
  F^{-1}(\partial_u\mathcal{C}) &\subset& \partial_u\mathcal{C},
\end{eqnarray}
so the image of the stable part of the partition boundary
$\partial_s\mathcal{C}$ is contained in $\partial_s\mathcal{C}$  and the
unstable part of the boundary $\partial_u\mathcal{C}$ contains its
preimage~\cite{GaspardBook}.  Such a~partition generates a~symbolic
dynamics with a $K$-letter alphabet 
which is a~topological Markov chain.  In the following we will
concentrate on the system~(\ref{eq:BNdef}), since most of the results
below can be translated directly to system~(\ref{eq:BSdef}), and we will
only emphasize important differences.

For system~(\ref{eq:BNdef}) we are able to find a Markov partition for
several values of the coupling parameter $\Delta$.  For example, for
$\Delta=1/2^k$ with natural $k$ this
partition is defined by a~set of horizontal lines $y_i=1-1/2^i$ where
$i=1,\ldots,k$ and a~vertical line at $x=1/2$.  It is easy to verify
that under the action of~(\ref{eq:BNdef}) each coordinate $y_i$ is
mapped to $y_{i-1}$ and eventually tends to $y_1=1/2$ which is mapped%
  \footnote{The system~(\ref{eq:BNdef}) is not continuous, so a~part of
  the neighbourhood of $y=1/2$ is mapped to $y=0$ and the remaining part
  of it to $y=1$ --- in our case both lines belong to the border of the
  Markov partition.}
to $y=0$.

Basing on this partition we can construct a~transition matrix
$\mathbf{T}$ with entries $T_{ij}$ equal to probabilities that the
system passes from the cell $i$ to the cell $j$.  For simple piecewise
linear maps an agreement between the resonance of the corresponding
Frobenius--Perron operator with the second largest modulus and the
eigenvalue of this matrix was observed~\cite{ChaosBook}.  The
dimension of the transition matrix for the system~(\ref{eq:BNdef}) with
$\Delta=1/2^k$ grows linearly with $k$ according to
$K=\dim\mathbf{T}=2(k+1)$, and one can easily obtain its eigenvalues.  For
instance, for $k=3$ ($\Delta=1/8$) the second largest eigenvalue is
equal $\lambda_2^+\approx-0.8846$, while in case of negative
coupling~(\ref{eq:BSdef}) $\lambda_2^- = -\lambda_2^+$ --- see
Appendix~\ref{app:Tmat}.

The transition matrix $\mathbf{T}$ for the system~(\ref{eq:BNdef})
enables us to specify periodic orbits of the system.
The number of periodic points is given directly by
\begin{equation}\label{eq:pp_trT}
  |\{(x,y):(x,y)=F^n(x,y)\}| = \tr \mathsf{T}^n\ ,
\end{equation}
where we introduced the connectivity matrix $\mathsf{T}$ (sometimes
called topological transition matrix~\cite{GaspardBook}) defined as
a~transition matrix with all non-zero entries replaced by 1,
\begin{equation}
  \mathsf{T}_{ij} = \left\{\begin{array}{ll}
    1 & \mathrm{if}\ \mathbf{T}_{ij}>0\\
    0 & \mathrm{otherwise}\end{array}\right.\ .
\end{equation}
It is worth to note that formula~(\ref{eq:pp_trT}) is valid only if
none of the periodic orbit crosses the boundary of the partition ---
otherwise we have to take into account that a~given symbolic sequence
may not define a~point in the phase space uniquely, so it might happen
that one orbit is counted more than once.  One can verify that there is
no such problem for system~(\ref{eq:BNdef}), but this is not the case
for system~(\ref{eq:BSdef}).  Periodic sequences $\omega_1=B_1A_2$ and
$\omega_2=B_2A_1$ (see Fig.~\ref{fig:BSdef}) correspond to the orbits
starting from the same initial point $x=1/2$, $y=1/3$ which belongs to
the partition line.  This fact needs to be taken into account during
calculations of the traces of the FP operator.

Note also that the systems~(\ref{eq:BNdef}) and~(\ref{eq:BSdef}),
originally defined on the square (torus) may also represent dynamics on
the sphere, where $y \rightarrow \cos\theta$ and $x \rightarrow
\varphi$.  In this case the entire line $y=0$ has to be identified with
the south pole (and line $y=1$ with the north pole, respectively), so
the number of periodic orbits in both systems may differ.

\subsection{Approximation of the Frobenius--Perron spectrum}
\label{sec:noise}

We are not able to find analytically resonances of the
systems~(\ref{eq:BNdef}) and (\ref{eq:BSdef}) (apart from the second
largest which we obtain from the transition matrix).  To approximate
them we follow an approach developed
in~\cite{OPSZ2000,fishman,weber1,OZ2001,weber2} and introduce
a~stochastic perturbation into the system.  This allows us to represent
the Frobenius--Perron operator corresponding to the system with noise as
a~finite dimensional matrix.

\begin{figure}
  \centering
  \includegraphics[width=.6\textwidth]{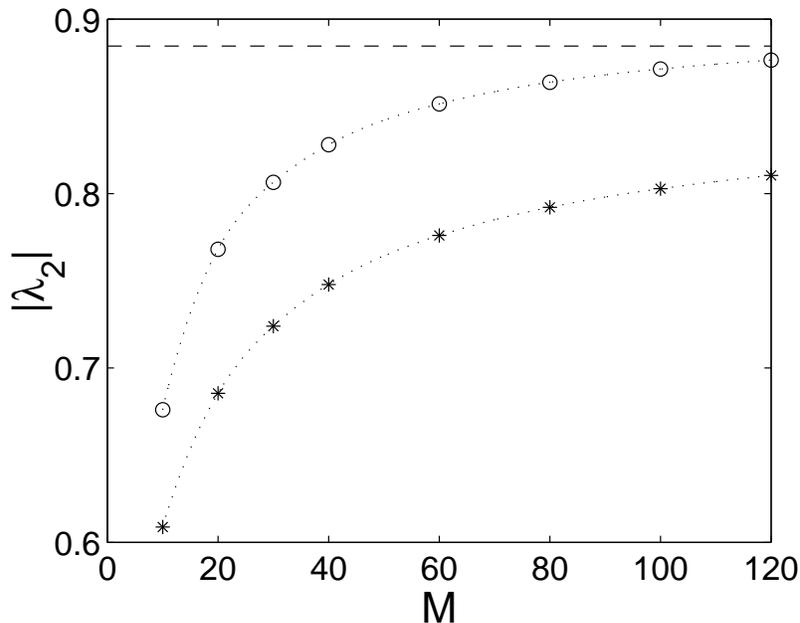}
  \caption{Dependence of the modulus of the second largest eigenvalue of
  the FP operator for the stochastically perturbed coupled baker maps on
  the sphere on the noise parameter $M$.  Values for the
  system~(\ref{eq:BNdef}) and~(\ref{eq:BSdef}) are represented by
  $\circ$ and $\ast$, respectively (in both cases $\Delta=1/8$).
  The dashed line represents the deterministic limit.}
  \label{fig:B2eig}
\end{figure}

\begin{figure}
  \psset{yunit=2.7\psunit,xunit=5.4\psunit,hatchsep=6pt}
  \begin{pspicture}(-.1,-.15)(1.1,1.1)
    \small
    \psline{<->}(0,1.1)(0,0)(1.1,0)
    \rput[r](-.05,.5){$t$} \rput[r](-.03,0){$-1$} \rput[r](-.03,1){$1$}
    \rput[t](.5,-.1){$\varphi$} \rput[t](0,-.1){$0$} \rput[t](1,-.1){$2\pi$}
    \psset{fillstyle=hlines,hatchangle=75,linestyle=none}
    \psframe(0,0)(.375,1) \psframe(.5,0)(.625,1)
    \psset{fillstyle=none,linestyle=solid}
    \multips(.125,0)(.125,0){7}{\psline(0,0)(0,1)}
    \psframe(0,0)(1,1) \psline(.25,.5)(.75,.5)
    \psset{linewidth=2.3pt}
    \psline(0,0)(0,1) \psline(.375,0)(.375,1)
    \psline(.5,0)(.5,1) \psline(.625,0)(.625,1)
    \psset{linecolor=white}
    \qdisk(.0625,.5){0.25}\rput(.0625,.5){$\mathbf{C_1''}$} \rput(.8125,.5){$\mathbf{B_2''}$}
    \qdisk(.1875,.5){0.25}\rput(.1875,.5){$\mathbf{B_1''}$} \rput(.9375,.5){$\mathbf{C_2''}$}
    \qdisk(.3125,.75){0.25}\rput(.3125,.75){$\mathbf{A_1''}$} \rput(.4375,.75){$\mathbf{C_2''}$}
    \qdisk(.3125,.25){0.25}\rput(.3125,.25){$\mathbf{C_1''}$} \rput(.4375,.25){$\mathbf{A_2''}$}
    \qdisk(.5625,.75){0.25}\rput(.5625,.75){$\mathbf{C_1''}$} \rput(.6875,.75){$\mathbf{A_2''}$}
    \qdisk(.5625,.25){0.25}\rput(.5625,.25){$\mathbf{A_1''}$} \rput(.6875,.25){$\mathbf{C_2''}$}
  \end{pspicture}\hfill
  \begin{pspicture}(-.1,-.15)(1.1,1.1)
    \small
    \psline{<->}(0,1.1)(0,0)(1.1,0)
    \rput[r](-.05,.5){$t$} \rput[r](-.03,0){$-1$} \rput[r](-.03,1){$1$}
    \rput[t](.5,-.1){$\varphi$} \rput[t](0,-.1){$0$} \rput[t](1,-.1){$2\pi$}
    \psset{fillstyle=hlines,hatchangle=75,linestyle=none}
    \psframe(0,0)(.125,1) \psframe(.25,0)(.5,1) \psframe(0.75,0)(.875,1)
    \psset{fillstyle=none,linestyle=solid}
    \multips(.125,0)(.125,0){7}{\psline(0,0)(0,1)}
    \psframe(0,0)(1,1) \psline(0,.5)(.25,.5) \psline(.75,.5)(1,.5)
    \psset{linewidth=2.3pt}
    \psline(0,0)(0,1) \psline(.125,0)(.125,1) \psline(.25,0)(.25,1)
    \psline(.5,0)(.5,1) \psline(.75,0)(.75,1) \psline(.875,0)(.875,1)
    \psset{linecolor=white}
    \qdisk(.0625,.75){0.25}\rput(.0625,.75){$\mathbf{A_1''}$} \rput(.1875,.75){$\mathbf{C_2''}$}
    \qdisk(.0625,.25){0.25}\rput(.0625,.25){$\mathbf{C_1''}$} \rput(.1875,.25){$\mathbf{A_2''}$}
    \qdisk(.3125,.5){0.25}\rput(.3125,.5){$\mathbf{A_1''}$}  \rput(.5625,.5){$\mathbf{B_2''}$}
    \qdisk(.4375,.5){0.25}\rput(.4375,.5){$\mathbf{B_1''}$}  \rput(.6875,.5){$\mathbf{A_2''}$}
    \qdisk(.8125,.75){0.25}\rput(.8125,.75){$\mathbf{C_1''}$} \rput(.9375,.75){$\mathbf{A_2''}$}
    \qdisk(.8125,.25){0.25}\rput(.8125,.25){$\mathbf{A_1''}$} \rput(.9375,.25){$\mathbf{C_2''}$}
  \end{pspicture}
  \caption{Second iteration of the systems~(\ref{eq:BNdef}) (left)
  and~(\ref{eq:BSdef}) (right).}
  \label{fig:sndstep}
\end{figure}
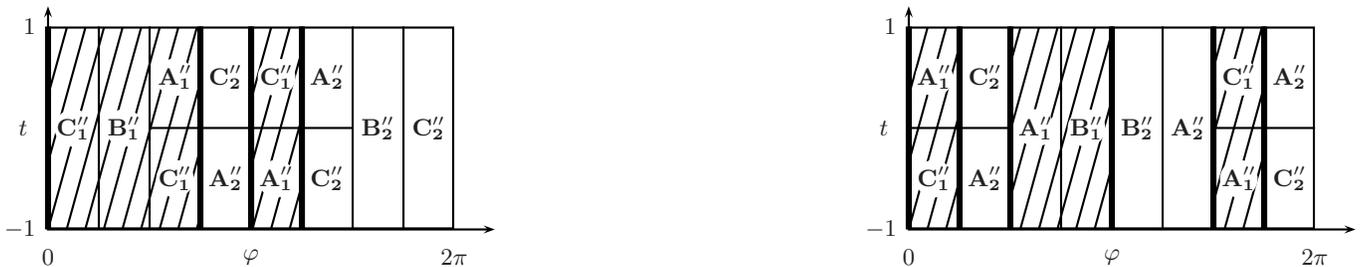

In Section~\ref{sec:quant} we define quantum propagators corresponding
to systems~(\ref{eq:BNdef}) and~(\ref{eq:BSdef}). 
In order to use well known  SU(2) coherent
states~\cite{Radcliffe,Arecchi,Glauber,Perelomov} we convert appropriate
definitions of the classical systems into the unit sphere by applying 
the Mercator projection. 
More formally, we replace $x$ with $\varphi=2\pi x$ and $y$ with
$t=\cos\theta=2y-1$, where $\varphi$ and $\theta$
are the usual spherical coordinates.  
The parameter $\Delta\in (0,\frac{1}{2})$ defined previously 
plays the same role of the coupling constant. 
Accordingly we
should approximate resonances for the classical system defined on the sphere.
A~possible choice of the probability density of the stochastic
perturbation which transforms a~point $\Omega=(\theta,\varphi)$ on the
sphere into $\Omega'=(\theta',\varphi')$ is
\begin{eqnarray}
  \lefteqn{\mathcal{P}_M(\Omega',\Omega) = } && \nonumber\\
  & = & \frac{M+1}{4\pi}\left(\frac{1+\cos\Xi}{2}\right)^M
    \label{eq:noise}\\
  & = & \frac{M+1}{4\pi}\sum_{k,l=0}^M {M\choose k}{M\choose l}
    (\sin^{k+l}\frac\theta2\cos^{2M-(k+l)}\frac\theta2 {\rm e}^{{\rm i}(l-k)\varphi})
    \times\nonumber\\
  && \times(\sin^{k+l}\frac{\theta'}2\cos^{2M-(k+l)}\frac{\theta'}2
    {\rm e}^{-{\rm i}(l-k)\varphi'}) \nonumber \ ,
\end{eqnarray}
where $M$ is an arbitrary natural number. 
Here $\Xi$ is the angle formed by two vectors pointing toward the
points $\Omega$ and $\Omega'$ on the sphere.  As discussed
in~\cite{OPSZ2000,OZ2001} the matrix representation of the
Frobenius--Perron operator for system with such a~noise is finite
dimensional --- the last equation in~(\ref{eq:noise}) helps to identify
the basis functions of the matrix representation.  For the probability
distribution~(\ref{eq:noise}) the dimension of the matrix is
$(M+1)^2$ and grows to infinity in the deterministic limit $M
\rightarrow \infty$, for which $\mathcal{P}_\infty(\Omega',\Omega)
= \delta(\Omega-\Omega')$ with respect to the uniform measure on the sphere,
${\rm d}\Omega=\sin\theta{\rm d}\theta\,{\rm d}\varphi$.  
However, for any finite $M$
one obtains a~finite representation of the FP operator and may
diagonalize it numerically%
  \footnote{Confined by the computer resources available we could
  investigate numerically the systems with the noise parameter
  $M\leq120$.}.
In this way we obtain the exact spectrum for the system with noise and
by decreasing the noise amplitude we can approximate the resonances of
the deterministic system.  Fig.~\ref{fig:B2eig} presents the dependence
of the modulus of the second largest eigenvalue $\lambda_2$ of
systems~(\ref{eq:BNdef}) and~(\ref{eq:BSdef}) subjected to additive
noise~(\ref{eq:noise}) on the noise parameter~$M$%
  \footnote{The phase of $\lambda_2$ for systems (\ref{eq:BNdef})
  and~(\ref{eq:BSdef}) does not depend on the noise parameter $M$ and is
  equal $0$ and $\pi$, respectively.}.
The deterministic limit --- represented in
Fig.~\ref{fig:B2eig} as a~dashed line --- is the same for both systems
and is obtained from the transition matrix defined in the previous
section.
The most striking observation is that although the deterministic value of
$|\lambda_2|$ for both systems is the same, the approximations of
$\lambda_2$ for a~given value of the noise parameter $M$ differ.  This can
be understood with the help of Fig.~\ref{fig:sndstep}, where we showed
the result of the second iteration of the corresponding systems.
The total length of the boundary between subsystem~$1$ and~$2$ for the
negative coupling (right plot) is one and half times larger than for the
positive coupling.  Thus in the case of the negative coupling it is more
likely that points will move from one subsystem to the other one under
the action of the stochastic perturbation.  The overall decay rate in
the presence of the noise is thus faster than in the case of the
positive coupling which is reflected in the spectrum as a~smaller
modulus of the subleading eigenvalue $\lambda_2$.

\subsection{Traces of the Frobenius--Perron operator}\label{sec:traces}

For the semiclassical analysis we will need the traces of the
Frobenius--Perron operator $P$ associated with the classical system.
Approximation of the traces with the help of stochastic perturbation is
straightforward, since we obtain eigenvalues $\lambda_i$ of the FP
operator for the noisy system and we can calculate the traces directly
from the definition
\begin{equation}\label{eq:trres}
  \tr P^n = \sum_i \lambda_i^n\ .
\end{equation}
In order to calculate the traces with the use of periodic orbits we will
use the integral representation of the FP operator~(\ref{eq:FPdef}).  To
obtain the expression for the traces it is sufficient to identify
initial and final points in this expression, that is
\begin{equation}\label{eq:trace}
  \tr P^n = \int {\rm d}\Omega\, \delta(\Omega-F^n(\Omega))
    = \sum_{\Omega_i:\Omega_i=F^n(\Omega_i)}
      \frac1{|\det(1-J^n(\Omega_i))|}\ ,
\end{equation}
where $J^n(\Omega_i)$ denotes the Jacobian matrix of the mapping $F^n$
evaluated in point $\Omega_i$ and we make use of the properties of the
$\delta$ function in the last equality.  The calculation of the
denominator in~(\ref{eq:trace}) is easy, since our systems are purely
hyperbolic with constant stretching and squeezing by a~factor of two.
Thus the contribution of each fixpoint to the trace of $P^n$ is equal to
$[2^n+2^{-n}-2]^{-1}$.  The only periodic points which need special
attention are the south and the north pole since --- due to the
discontinuities --- expression~(\ref{eq:trace}) is not well defined in
these points.  In order to calculate the contribution from these points
we ``regularized'' the integral~(\ref{eq:trace}) --- see
Appendix~\ref{app:poles}.
\begin{figure}
  \centering
  \includegraphics[width=.5\textwidth]{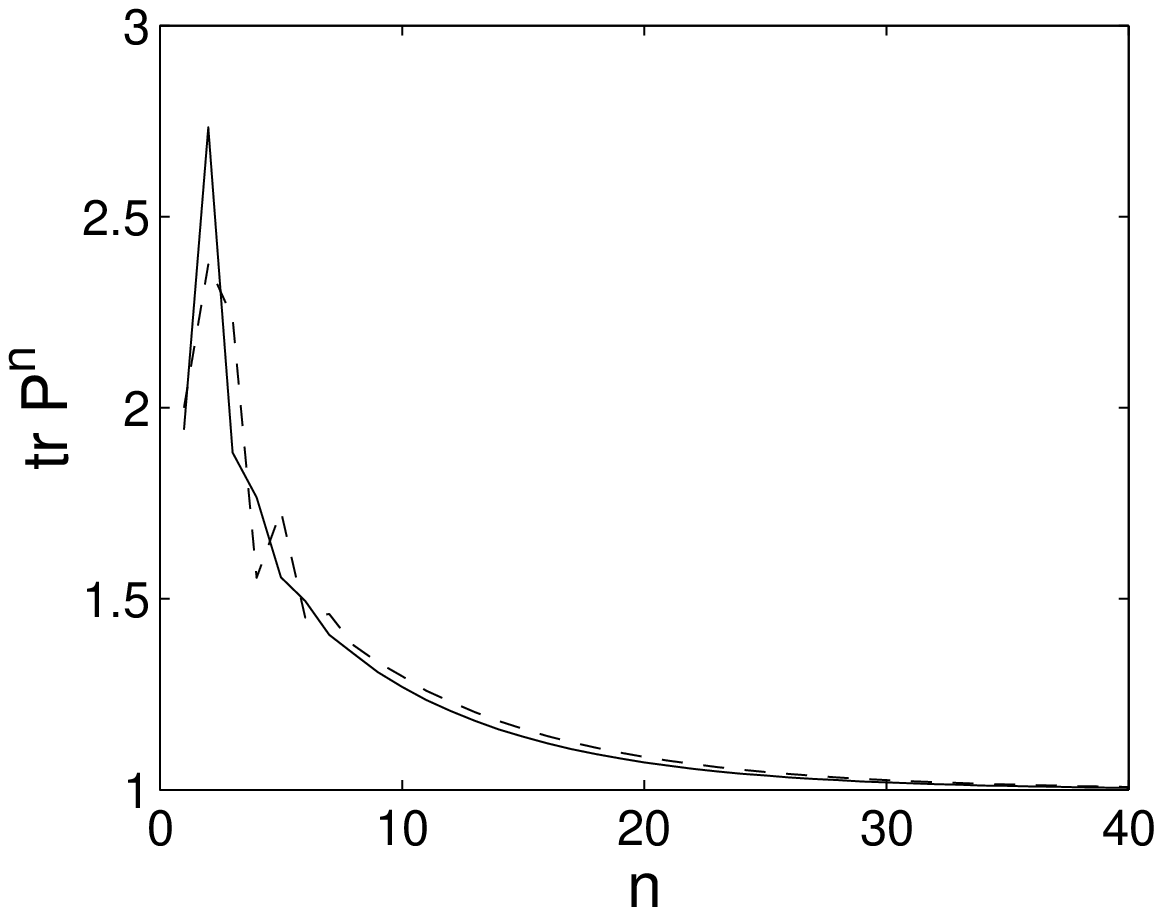}%
    \rput(-.05\textwidth,.33\textwidth){(a)}%
  \includegraphics[width=.5\textwidth]{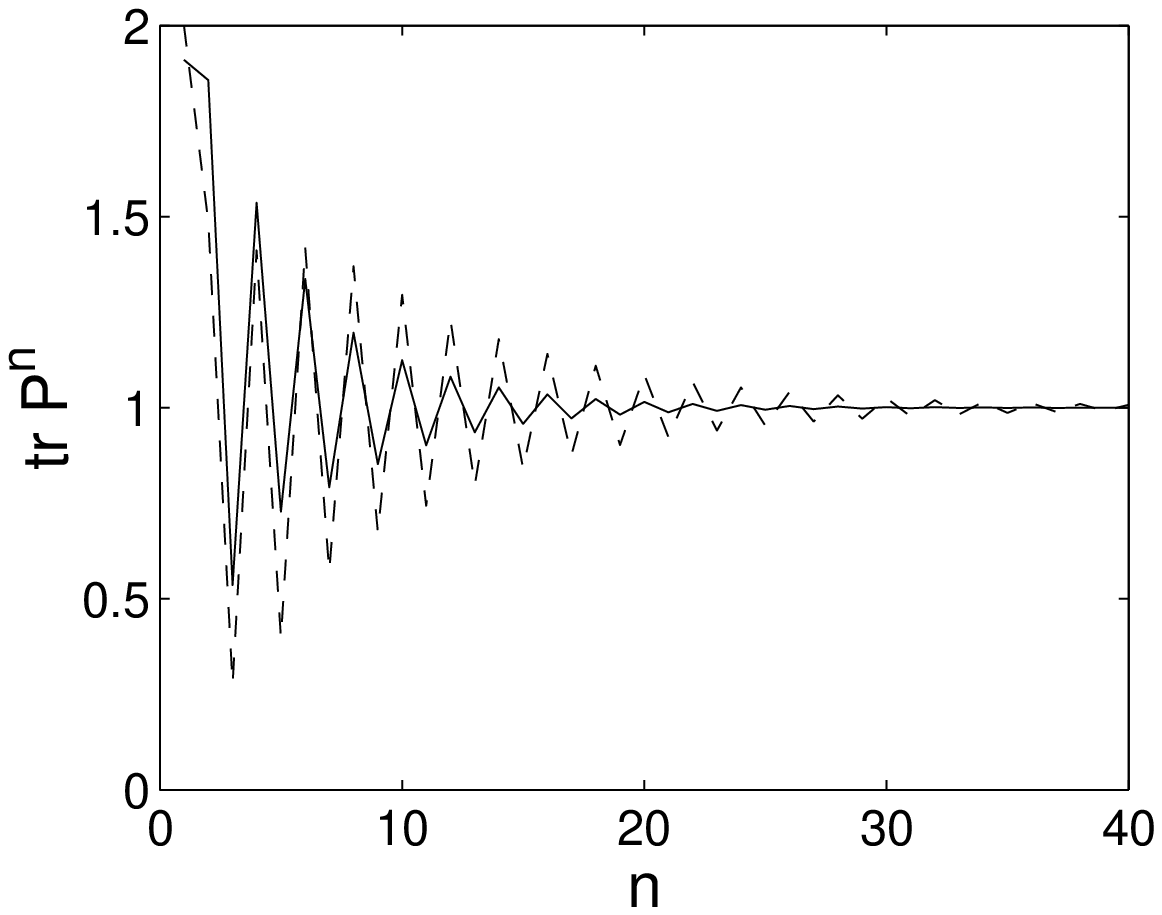}%
    \rput(-.05\textwidth,.33\textwidth){(b)}\\
  \includegraphics[width=.5\textwidth]{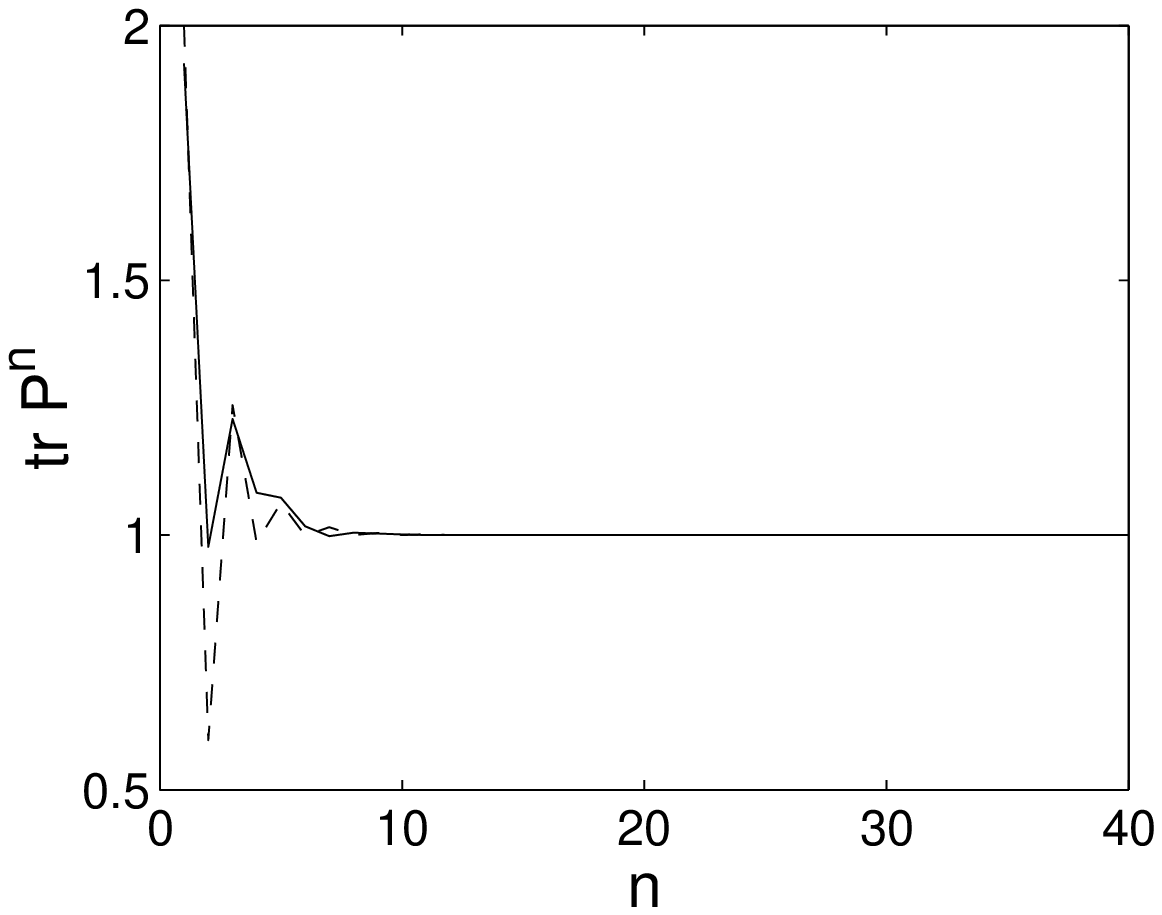}%
    \rput(-.05\textwidth,.33\textwidth){(c)}
  \caption{Traces of the FP operator: (a) positive
  coupling~(\ref{eq:BNdef}) with $\Delta=1/8$, (b) negative
  coupling~(\ref{eq:BSdef}) ($\Delta=1/8$) and (c) full coupling
  corresponding to the system (\ref{eq:BNdef}) with $\Delta=1/2$ (or
  to (\ref{eq:BSdef}) also with $\Delta=1/2$).  Solid lines represent
  traces calculated from the approximate spectrum of the FP operator
  while the traces obtained from the periodic orbits are plotted with
  dashed lines.}
  \label{fig:traces}
\end{figure}
Having done that we can combine the contribution of the periodic points
with their numbers calculated from~(\ref{eq:pp_trT}). 
The traces, Fig.~\ref{fig:traces}, calculated
from the approximate spectrum and periodic orbits are represented by
solid and dashed lines, respectively.  These numerical results
demonstrate good agreement between two different methods of computing
the traces of the FP operator.  Discrepancies visible in
Fig.~\ref{fig:traces}b and \ref{fig:traces}c are due to the fact that
for the system~(\ref{eq:BSdef}) the effective coupling is stronger as
discussed at the end of Sec.~\ref{sec:noise}.

\section{Quantum propagator}
\label{sec:quant}

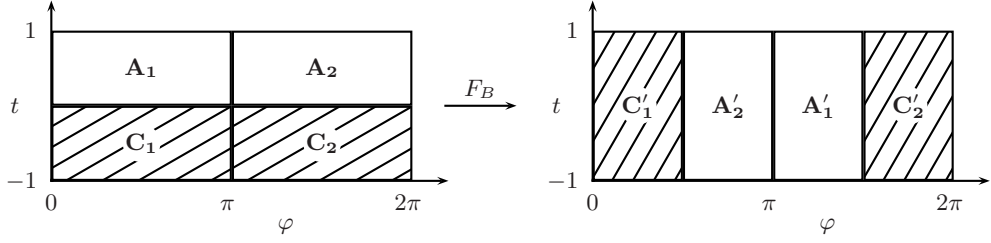
\begin{figure}
  \centering
  \begin{pspicture}(-.5,-1.7)(12.5,1.3)
    \renewcommand\pshlabel[1]{\ifnum #1=0 $0$%
        \else \ifnum #1=1 $\pi$ \else $#1\pi$ \fi \fi}
    \psset{xunit=2.4cm,hatchsep=6pt}
    \rput(0,0){
      \psaxes[ticks=none,Oy=-1,Dy=2]{->}(0,-1)(2.2,1.4)
        \rput(-.2,0){$t$} \rput(1.3,-1.6){$\varphi$}
      \psset{fillstyle=none}
      \psframe(0,0)(1,1) \psframe(1,0)(2,1)
        \rput(.5,.5){$\mathbf{A_1}$} \rput(1.5,.5){$\mathbf{A_2}$}
      \psset{hatchangle=30,fillstyle=hlines}
      \psframe(0,-1)(1,0) \psframe(1,-1)(2,0)
	\psset{linecolor=white}
        \qdisk(.5,-.5){0.25}\rput(.5,-.5){$\mathbf{C_1}$}
	\qdisk(1.5,-.5){0.25}\rput(1.5,-.5){$\mathbf{C_2}$}
    }
    \psline{->}(2.2,0)(2.6,0) \rput[b](2.4,.1){$F_B$}
    \rput(3,0){
      \psaxes[ticks=none,Oy=-1,Dy=2]{->}(0,-1)(2.2,1.4)
        \rput(-.2,0){$t$} \rput(1.3,-1.6){$\varphi$}
      \psset{fillstyle=none}
      \psframe(.5,-1)(1,1) \psframe(1,-1)(1.5,1)
        \rput(.75,0){$\mathbf{A'_2}$} \rput(1.25,0){$\mathbf{A'_1}$}
      \psset{fillstyle=hlines,hatchangle=60}
      \psframe(0,-1)(.5,1) \psframe(1.5,-1)(2,1)
	\psset{linecolor=white}
        \qdisk(.25,0){0.25}\rput(.25,0){$\mathbf{C'_1}$}
	\qdisk(1.75,0){0.25}\rput(1.75,0){$\mathbf{C'_2}$}
    }
  \end{pspicture}
  \caption{Baker map on the sphere in Mercator projection
  $(t=\cos\theta,\varphi)$.  This map is the full coupling limit
  ($\Delta=1/2$) of the systems (\ref{eq:BNdef}) and (\ref{eq:BSdef}).}
  \label{fig:CSBaker}
\end{figure}

In the construction of the quantum propagator corresponding to the
systems~(\ref{eq:BNdef}) and~(\ref{eq:BSdef}) we rely on the results
presented in~\cite{Pakonski1999} in which the quantum baker map was
defined on the sphere.  The corresponding classical system is obtained
from (\ref{eq:BNdef}) and (\ref{eq:BSdef}) in the full coupling limit
$\Delta=1/2$ and its action is visualized in Fig.~\ref{fig:CSBaker}.
The construction of the quantum system is based on the matrix
representation of the rotation around the $y$--axis by an angle
of~$\pi/2$
\begin{equation}
  R_{m',m} = \langle j,m|{\rm e}^{-{\rm i}\frac\pi2\hat J_y}|j,m'\rangle\ ,
\end{equation}
where $\hat J_i$ is the $i$-th component of the angular momentum
operator $\hat J$ and $|j,m\rangle$ is an eigenvector of the $\hat J_z$
operator, $\hat J_z |j,m\rangle = m |j,m\rangle$ with $m=-j,\ldots,j$.  In the
following we choose half--integer spin values $j$, so the size of the
rotation matrix $N=2j+1$ is even.  The resulting quantum propagator is
defined as~\cite{Pakonski1999}
\begin{equation}\label{eq:Bprop}
  \hat U_B = R^{-1}\left[\begin{array}{cc}R' & 0\\ 0 & R''\end{array}\right]\ ,
\end{equation}
where $R'$ and $R''$ are matrices created by taking halves
(respectively upper and lower) of every second column of the rescaled
Wigner rotation matrix $R$
\begin{eqnarray}
  R'_{k,m} &=& \sqrt2\,R_{k,2m-j}\qquad k,m=\frac12, \ldots , j\ ,\\
  R''_{k,m} &=& \sqrt2\,R_{k,2m+j}\qquad k,m=-j, \ldots , -\frac12\ .
\end{eqnarray}
The construction~(\ref{eq:Bprop}) is similar to the original quantum map
on the torus proposed by Balazs and Voros~\cite{BV}: instead of the
Fourier matrix we use the Wigner rotation matrix $R$.

Now it is crucial to note that if we additionally exchange the parts
$A_1'$ and $A_2'$ in Fig.~(\ref{fig:CSBaker}), we obtain the
uncoupled version of the map~(\ref{eq:BNdef}) for $\Delta=0$.  The same
happens if we exchange the parts $C_1'$ and $C_2'$ --- we obtain then
map~(\ref{eq:BSdef}) for $\Delta=0$.  On the other hand a~partial
exchange of these regions will result in maps with $\Delta>0$.

\begin{figure}
  \centering
  \psset{yunit=1.9\psunit,xunit=3.6\psunit}
  \begin{pspicture}(0,0)(3.65,2.8)
  \rput[lb](0,1.1){(b)}
  \rput[lb](0,0){
    \psframe(0,.5)(.5,1) \psframe(0,.2)(.5,.5) \psframe(0,0)(.5,.2)
      \rput(.25,.75){$A_1$} \rput(.25,.35){\scriptsize$B_1$}
      \rput(.25,.1){\scriptsize$C_1$}
    \psframe(.5,.5)(1,1) \psframe(.5,.2)(1,.5) \psframe(.5,0)(1,.2)
      \rput(.75,.75){$A_2$} \rput(.75,.35){\scriptsize$B_2$}
      \rput(.75,.1){\scriptsize$C_2$}
  }
  \psline{->}(1.05,.5)(1.3,.5) \rput[b](1.175,.54){\small$R_S(\Delta)$}
  \rput[lb](1.35,0){
    \psframe(0,.5)(.5,1) \psframe(0,.2)(.5,.5) \psframe(0,0)(.5,.2)
      \rput(.25,.75){$A_1$} \rput(.25,.35){\scriptsize$B_2$}
      \rput(.25,.1){\scriptsize$C_1$}
    \psframe(.5,.5)(1,1) \psframe(.5,.2)(1,.5) \psframe(.5,0)(1,.2)
      \rput(.75,.75){$A_2$} \rput(.75,.35){\scriptsize$B_1$}
      \rput(.75,.1){\scriptsize$C_2$}
  }
  \psline{->}(2.4,.5)(2.6,.5) \rput[b](2.5,.54){$F_B$}
  \rput[lb](2.65,0){
    \psframe(.75,0)(1,.4) \psframe(0,.4)(.25,1) \psframe(.25,0)(.5,1)
      \rput(.875,.2){$C'_2$} \rput(.125,.7){$B'_2$} \rput(.375,.5){$A'_2$}
    \psframe(0,0)(.25,.4) \psframe(.75,.4)(1,1) \psframe(.5,0)(.75,1)
      \rput(.125,.2){$C'_1$} \rput(.875,.7){$B'_1$} \rput(.625,.5){$A'_1$}
  }
  \rput[lb](0,1.5){
    \psframe(0,.8)(.5,1) \psframe(0,.5)(.5,.8) \psframe(0,0)(.5,.5)
      \rput(.25,.9){\scriptsize$A_1$} \rput(.25,.65){\scriptsize$B_1$}
      \rput(.25,.25){$C_1$}
    \psframe(.5,.8)(1,1) \psframe(.5,.5)(1,.8) \psframe(.5,0)(1,.5)
      \rput(.75,.9){\scriptsize$A_2$} \rput(.75,.65){\scriptsize$B_2$}
      \rput(.75,.25){$C_2$}
  }
  \psline{->}(1.05,2)(1.3,2) \rput[b](1.175,2.04){\small$R_N(\Delta)$}
  \rput[lb](1.35,1.5){
    \psframe(0,.8)(.5,1) \psframe(0,.5)(.5,.8) \psframe(0,0)(.5,.5)
      \rput(.25,.9){\scriptsize$A_1$} \rput(.25,.65){\scriptsize$B_2$}
      \rput(.25,.25){$C_1$}
    \psframe(.5,.8)(1,1) \psframe(.5,.5)(1,.8) \psframe(.5,0)(1,.5)
      \rput(.75,.9){\scriptsize$A_2$} \rput(.75,.65){\scriptsize$B_1$}
      \rput(.75,.25){$C_2$}
  }
  \psline{->}(2.4,2)(2.6,2) \rput[b](2.5,2.04){$F_B$}
  \rput[lb](2.65,1.5){
    \psframe(0,0)(.25,1) \psframe(.5,.6)(.75,1) \psframe(.25,0)(.5,.6)
      \rput(.125,.5){$C'_1$} \rput(.625,.8){$A'_1$} \rput(.375,.3){$B'_1$}
    \psframe(.75,0)(1,1) \psframe(.25,.6)(.5,1) \psframe(.5,0)(.75,.6)
      \rput(.875,.5){$C'_2$} \rput(.375,.8){$A'_2$} \rput(.625,.3){$B'_2$}
  }
  \rput[lb](0,2.6){(a)}
  \end{pspicture}
  \caption{Construction of the maps~(\ref{eq:BNdef})~(a)
  and~(\ref{eq:BSdef})~(b) with use of the baker map on the sphere
  $F_B$~(cf.~Fig.~\ref{fig:CSBaker}).  Exchange of the cells
  $B_1\leftrightarrow B_2$ is denoted by~$R_N(\Delta)$ and~$R_S(\Delta)$
  respectively.}
  \label{fig:CB_to_BNS}
\end{figure}
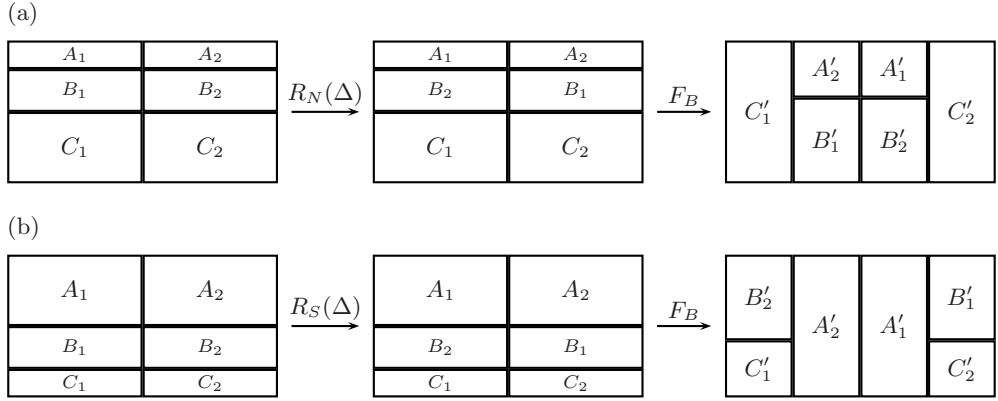

The only question left is how to modify the definition of the quantum
propagator in order to obtain the above mentioned exchange.  It is
sufficient to swap the cell $A_1$ with $A_2$ in Fig.~\ref{fig:CSBaker}
(or $C_1$ with $C_2$), or their parts, before applying the operator
$\hat U_B$ from~(\ref{eq:Bprop}).  Such an exchange may be
accomplished by a~rotation of the part of northern or southern
hemisphere (for map~(\ref{eq:BNdef}) and~(\ref{eq:BSdef}), respectively)
around the $z$--axis by angle $\pi$.  This procedure is presented in
Fig.~\ref{fig:CB_to_BNS} where this partial rotation is denoted by
$R_N(\Delta)$ and $R_S(\Delta)$.
Quantum operators corresponding to these rotations have simple
representation in the $|j,m\rangle$ basis, since they are just multiplication
of the vector of coefficients by a~phase factor ${\rm e}^{-{\rm i}m\pi}$.
In both cases the matrix representation is diagonal and for the
``positive'' coupling~(\ref{eq:BNdef}) the diagonal elements are equal
\begin{equation}\label{eq:NDelta}
  \hat R_N(\Delta)_{k,k} = \left\{\begin{array}{lll}
      1 & \mathrm{for} & k < N\Delta \\
      {\rm e}^{-{\rm i}(j-k+1)\pi} & \mathrm{for} & k \in (N\Delta, N/2) \\
      1 & \mathrm{for} & k > N/2
    \end{array}\right.\ ,
\end{equation}
while for ``negative'' coupling~(\ref{eq:BSdef}) the appropriate rotation
operator has the following representation
\begin{equation}\label{eq:SDelta}
  \hat R_S(\Delta)_{k,k} = \left\{\begin{array}{lll}
      1 & \mathrm{for} & k < N/2 \\
      {\rm e}^{-{\rm i}(j-k+1)\pi} & \mathrm{for} & k \in (N/2, N(1-\Delta)) \\
      1 & \mathrm{for} & k > N(1-\Delta)
    \end{array}\right.\ ,
\end{equation}
where $N=2j+1$ is the dimension of the Hilbert space (we confine ourself
to such values of $N$ and $\Delta$ that $N\Delta$ is integer).  Using
these operators we define unitary quantum maps corresponding to
classical systems~(\ref{eq:BNdef}) and~(\ref{eq:BSdef}), respectively,
\begin{eqnarray}
  \hat U_N &=& \hat U_B\hat R_N(\Delta)\label{eq:hatBN}\\
  \hat U_S &=& \hat U_B\hat R_S(\Delta)\label{eq:hatBS}\ ,
\end{eqnarray}
with $\hat U_B$ defined by Eq.~(\ref{eq:Bprop}).

\subsection{Comparison of classical and quantum dynamics}

In order to illustrate correctness of the proposed construction of the
quantum propagators, we make use of periodic orbits of the classical
systems.  Suppose that for initial conditions for quantum dynamics we
choose a~wave function localized around some periodic point of
the classical system.  After the time equal to the period of the
classical orbit the probability of measuring the system near the initial
point in the phase space should be large.  More precisely, for the
initial state we choose the SU(2) coherent state $|\theta,\varphi\rangle$
localized in point $(\theta, \varphi)$.  In the angular momentum
representation a~coherent state can be generated by a~rotation of the
state $|j,m=j\rangle$, which fulfills the smallest uncertainty
relation~\cite{Arecchi,Glauber}, as $|\theta,\varphi\rangle
= R(\theta,\varphi)|j,j\rangle$.  Now we will investigate the so called
return probability
\begin{equation}\label{eq:retprob}
  |Q_{U^n}(\theta,\varphi)|^2 = |\langle\theta,\varphi|U^n|\theta,\varphi\rangle|^2\ ,
\end{equation}
where $U$ is the quantum propagator.  The $Q$ function of an operator
$A$ is defined as
\begin{equation}
  Q_A = \langle\theta, \varphi| A |\theta, \varphi\rangle\ .
\end{equation}
For density operators the $Q$ function is also called Husimi function.
\begin{figure}[p]
  \centering
  \includegraphics[width=0.9\textwidth]{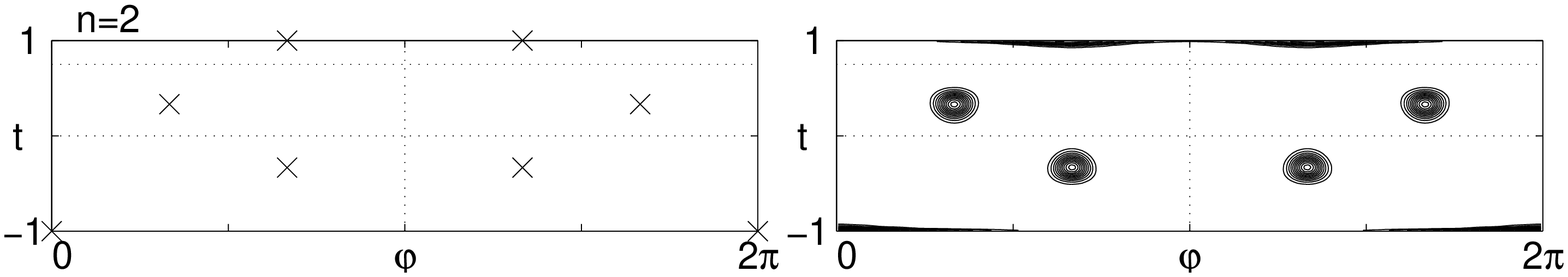}\\
  \includegraphics[width=0.9\textwidth]{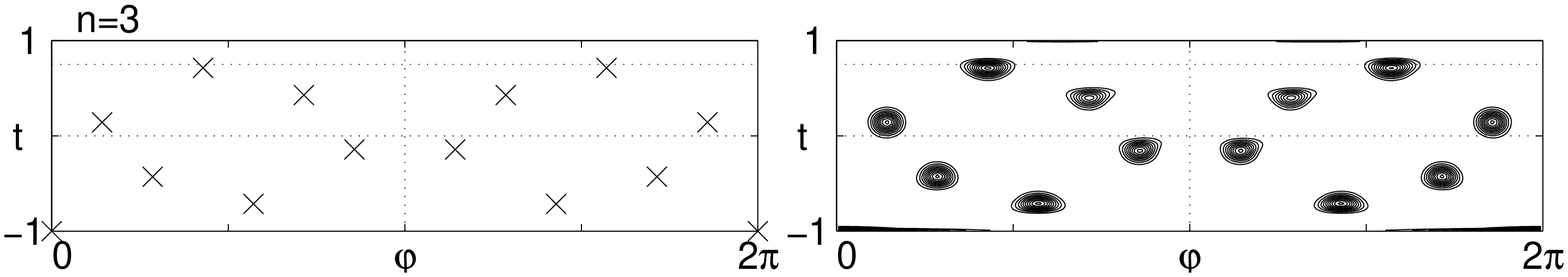}
  \makebox[\textwidth]{\hrulefill}\\[5mm]
  \includegraphics[width=0.9\textwidth]{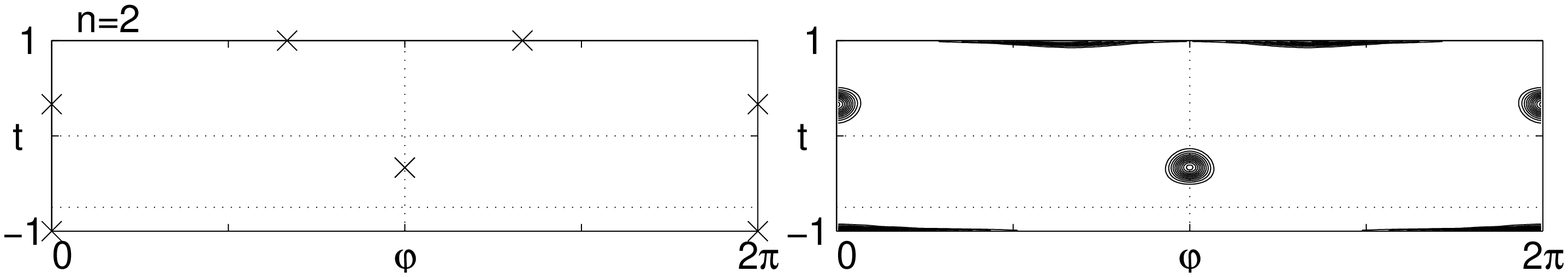}\\
  \includegraphics[width=0.9\textwidth]{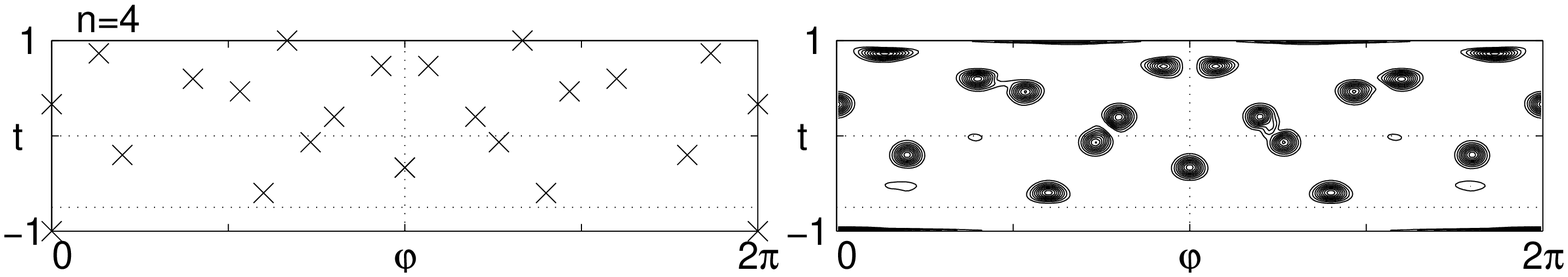}
  \caption{Classical periodic orbits (left hand side) for the
  systems~(\ref{eq:hatBN}) (upper part) and~(\ref{eq:hatBS}) (lower
  part) versus quantum return probability~(\ref{eq:retprob}) for
  selected lengths $n$ of the orbits ($\Delta=1/8$).}
  \label{fig:husimi}
\end{figure}
\begin{figure}[p]
  \centering
  \includegraphics[width=0.9\textwidth]{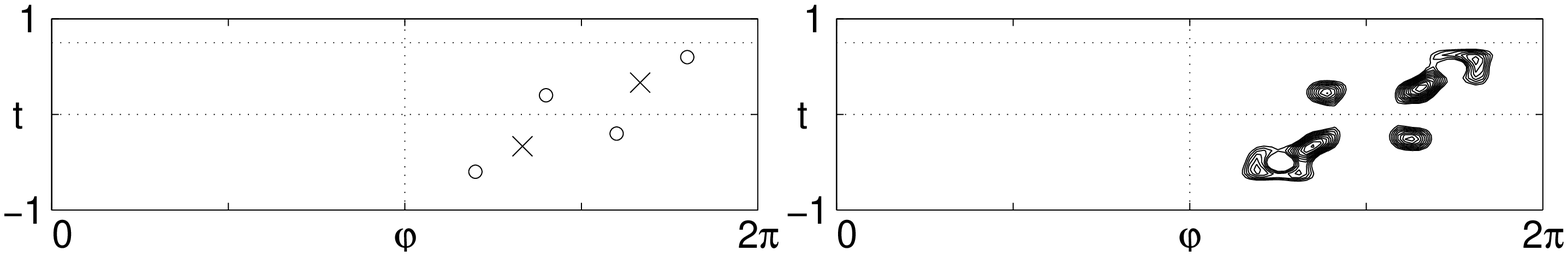}\\
  \includegraphics[width=0.9\textwidth]{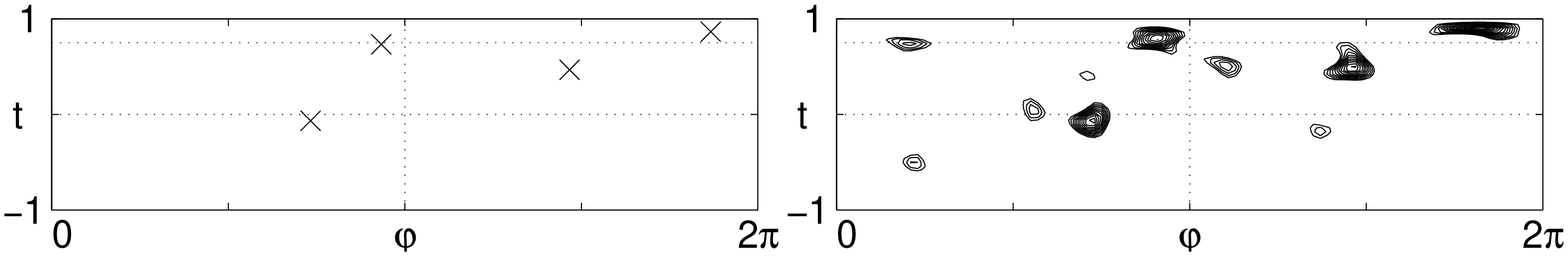}
  \caption{Periodic orbits of the classical system (left hand side)
  together with Husimi functions $Q_{|\psi\rangle\langle\psi|}$ of selected
  eigenvectors of quantum propagators showing corresponding scars.
  Upper part regards the model with positive coupling and shows that the
  selected
  eigenvector is scared by two orbits --- period--2 orbit denoted on left
  hand side by $\times$ and period 4 orbit denoted by $\circ$.  Lower
  part is for the system with negative coupling and in this case the
  selected eigenvector is scarred along period--4 orbit of the classical
  system.  In both cases $\Delta=1/8$.}
  \label{fig:scars}
\end{figure}
In a~vicinity of the
points $(\theta,\varphi)$, corresponding to periodic points of orbits
of length $n$, we may expect large values of the return probability.
The functions $|Q_{U^n}(\theta,\varphi)|^2$ for systems~(\ref{eq:hatBN})
and~(\ref{eq:hatBS}) are presented in Fig.~\ref{fig:husimi}: observe an
agreement between maxima of the quantum return
probability~(\ref{eq:retprob}) and the periodic points of the classical
system.

To emphasize the fact that regular periodic points, that is
those indicated by corresponding symbolic dynamics, have much more
influence than periodic points resulting from the boundary conditions we
only showed points obtained from the Markov partition --- without
corrections resulting from the topology of the system.  For example the
line $t=1$ is one point (north pole) so a periodic orbit of length two for
this value of $t$ coordinate is a~fixpoint of the map.  However, one
can see that quantum return probability for this point is much
higher for even iterations than for the odd ones.

One can also notice that some eigenvectors of the quantum propagator are
scarred~\cite{Heller,Heller1,Heller2,Kus,Stockmann} by classical
periodic orbits which is shown in Fig.~\ref{fig:scars}.
We conclude thus, that the procedure presented above indeed gives as
a~result quantum systems which correspond to~(\ref{eq:BNdef})
and~(\ref{eq:BSdef}).

\section{Averaged Overlaps of Husimi Eigenfunctions}\label{sec:overlaps}

We here discuss how Ruelle--Pollicott resonances rise 
deviations from random-matrix theory as an application 
of tailormade resonances. 
In \cite{chris} it has been shown that the classical resonances 
lead to semiclassical corrections of the localization of quantum eigenstates. 
In particular, it was shown that the probability of finding strongly 
phase-space overlapping quantum eigenfunctions increases if the difference 
of their (quasi-)eigenenergies is close to the phase of a classical resonance 
which corresponds to a slow correlation decay. 
On the other hand, if a pair of eigenfunctions strongly overlaps then each 
of them must be localized (scarred) in the same phase-space regions. 

In contrast to the numerical results of \cite{chris} which were obtained from a system 
with a classical mixed phase space we have here completely analytical classical 
results. Indeed, we do not calculate the resonances analytically, but for our 
purpose the traces of the Frobenius--Perron operator are sufficient. 
Here we briefly review those results that are relevant for the present discussion. 

We focus our investigation on pure-state Husimi functions of 
projectors of Floquet eigenstates, i.e. the phase-space representation 
of (quasi-)energy eigenstates, 
\begin{equation}
Q_{kk}\equiv Q_{|\phi_k\rangle\langle\phi_k|}=|\langle\theta,\varphi|\phi_k\rangle|^2
\,.
\end{equation}
Due to the normalization of a density operator, $\tr\rho=1$, the 
Husimi functions are $L^1$ normalized as 
\begin{equation}
\frac{N}{4\pi}
\int{\rm d}\Omega\,Q_\rho(\theta,\varphi)=1
\,.
\end{equation}

As a measure for phase-space localization we introduce the $L^2$ norm of a 
Husimi function, 
\begin{equation}
||Q_{kk}||^2=\int{\rm d}\Omega\,|\langle\theta,\varphi|\phi_k\rangle|^4
\,,
\end{equation}
which is the inverse participation ratio with respect to coherent states. 
Another property of interest is the phase-space overlap of two Husimi functions, 
\begin{equation}
||Q_{ik}||^2=\int{\rm d}\Omega\,|\langle\theta,\varphi|\phi_i\rangle|^2
|\langle\theta,\varphi|\phi_k\rangle|^2
\,.
\end{equation}
The notation $Q_{ik}$ is used, since the phase-space overlap is the same as 
the $L^2$ norm of the ``skew'' Husimi function $Q_{|\phi_i\rangle\langle\phi_k|}$. 
The physical meaning of the overlaps becomes obvious from Schwarz' inequality, 
\begin{equation}
||Q_{ik}||^2\le ||Q_{ii}||\,||Q_{kk}||
\,.
\label{schwarz}
\end{equation}
For large values of $||Q_{ik}||^2$ both Husimi functions 
must be localized in the same phase-space regions. 

The introduced measures, $L^2$ norm or phase-space overlap, 
prove amenable to semiclassical considerations. 
It might be obvious that the return probability becomes 
\begin{equation}
\frac{N}{4\pi}|\langle\theta,\varphi|U^n|\theta,\varphi\rangle|^2
\jinft\delta(\Omega-F^n(\Omega))
\end{equation}
in the classical limit. 
Integration over phase space leads to the trace of the Frobenius-Perron operator, 
\begin{equation}
\frac{N}{4\pi}\int{\rm d}\Omega\,|\langle\theta,\varphi|U^n|\theta,\varphi\rangle|^2
\jinft\tr P^n
\,.
\label{tracelim}
\end{equation}
Introducing the Floquet eigenstates on the left hand side one finds
\begin{eqnarray}
\frac{N}{4\pi}\int{\rm d}\Omega\,
|\langle\theta,\varphi|U^n|\theta,\varphi\rangle|^2&=&
\frac{N}{4\pi}\int{\rm d}\Omega\,
\left|\sum_k|\langle\theta,\varphi|\phi_k\rangle|^2{\rm e}^{-{\rm i}n\phi_k}\right|^2
\nonumber\\ &=&
\frac{N}{4\pi}\sum\limits_{ik}||Q_{ik}||^2\,{\rm e}^{-{\rm i}n(\phi_i-\phi_k)}
\,.
\end{eqnarray}
Fourier transformation of the latter expression leads to a sum of 
$\delta$ functions weighted by $L^2$ norms,
\begin{equation}
\sum\limits_{n=-\infty}^{\infty}\frac{{\rm e}^{{\rm i}n\omega}}{2\pi}
\sum\limits_{ik}||Q_{ik}||^2\,{\rm e}^{-{\rm i}n(\phi_i-\phi_k)}
=\sum\limits_{ik}||Q_{ik}||^2\,\delta(\omega-(\phi_i-\phi_k))
\label{kamm}
\,.
\end{equation}
For finite Hilbert-space dimension $N$ the relation (\ref{tracelim}) might be valid 
for finite times $|n|\le{\cal N}$. The truncated Fourier transform leads to a sum 
of smoothed $\delta$ functions, 
\begin{equation}
\overline{||Q_{ik}||^2}^{\Delta\omega}(\omega)
=\frac{4\pi}{N}
\sum_{n=-{\cal N}}^{\cal N} \tr P^n\frac{{\rm e}^{{\rm i}n\omega}}{2\pi}
\,.
\label{las}
\end{equation}
The stationary eigenvalue $1$ will be dropped; it would lead to a 
$\delta$ function after Fourier transformation in the limit ${\cal N}\rightarrow\infty$. 
To this end we identify the stationary eigenvalue in the Husimi representation. 
In \cite{chris} it has been shown that the stationary eigenvalue is related to 
the $L^1$ norms of Husimi functions. It turns out that the eigenvalue 1 can be dropped 
on the right hand side of (\ref{las}) if  one replaces the Husimi functions 
$Q_{kk}$ by $Q_{kk}'=Q_{kk}-\frac{1}{N}$ on the left hand side, 
where $\int{\rm d}\Omega\, Q_{kk}'=0$. 
(The prime will be dropped in the following.) 
For $n=0$ the integral on the left hand side of (\ref{tracelim}) gives the dimension $N$. 
We replace the traces by sums of the Ruelle-Pollicott resonances
(\ref{eq:trres}) 
and make use of the symmetry $\tr P^{-n}=\tr P^{n}$, 
\begin{equation}
\overline{||Q_{ik}||^2}^{\Delta\omega}(\omega)
=2\frac{N-1}{N}+\frac{4}{N}\sum_{n=1}^{\cal N}
\sum_\nu \lambda^n_\nu\cos n\omega 
\label{sumres}
\,.
\end{equation}
Note that the eigenvalue $1$ is also dropped in the leading order term. 
Assuming that the density of differences of Floquet eigenphases is almost 
constant, $N^2/2\pi$,  we finally get the result that the overlaps
averaged over an interval $\Delta\omega$ of differences of Floquet
eigenphases ($\omega_{ik}=\phi_i-\phi_k$) are given by traces of the
Frobenius--Perron operator, i.e. Ruelle--Pollicott resonances, 
\begin{eqnarray}
 \left\langle||Q_{ik}||^2\right\rangle(\omega)
&=&\frac{4\pi}{N^2}\left(
1-\frac{1}{N}+\frac{2}{N}\sum_{n=1}^{\infty}
({\rm tr}P^n-1)\cos n\omega\right)
\nonumber\\
&=&\frac{4\pi}{N^2}\left(
1-\frac{1}{N}+\frac{2}{N}\sum_\nu\sum_{n=1}^{\infty}
\lambda_\nu^n\cos n\omega\right)
\,.
\label{sccorr}
\end{eqnarray}
The constant term in parenthesis coincides up to the order $N^{-2}$ 
with the result of RMT \cite{chris}, 
$\left\langle||Q_{ik}||^2\right\rangle_{\rm RMT}=\frac{4\pi}{N(N+1)}$. 
The traces of the Frobenius--Perron operator are expanded in sums of the
Ruelle--Pollicott resonances, where a Fourier transform of each
resonance leads to a periodic Lorentzian distribution displaced by the
phase of the resonance. Note that the resonances 
are real or appear as complex conjugated pairs. 

The averaged phase-space overlaps (\ref{sccorr}) might be understood 
as a scar correlation function. Due to the Schwarz' inequality
(\ref{schwarz}) the probability to find a pair of scarred eigenfunctions
becomes large if the difference of their eigenenergies is close to the
phase of a resonance of large modulus, i.e. close to the position of
a strongly peaked Lorentzian. 

For numerical results 
we first have smoothed the sum of weighted $\delta$ functions 
by a convolution with a sinc function between its first zeros,  
\begin{equation}
\overline{||Q_{ik}||^2}^{\Delta\omega}(\omega)\propto
\int\limits_{\omega-\frac{\pi}{\cal N}}^{\omega+\frac{\pi}{\cal N}}
{\rm d}\omega'\frac{\sin ({\cal N}(\omega-\omega'))}
{\omega-\omega'}\sum_{ik}||Q_{ik}||^2\delta(\omega'-(\phi_i-\phi_k))
\,,
\label{smooth}
\end{equation}
where we have chosen ${\cal N}=10$. 
If one believes in validity of semiclassical methods up 
to the Ehrenfest time one may identify ${\cal N}$ as the Ehrenfest time. 
Anyway, ${\cal N}$ should be chosen that, on the one hand, quantum 
fluctuations are smoothed out, but on the other hand, the Lorentzian peaks 
keep their widths and heights. 
The averaged overlaps are entered by dividing the latter smoothed function 
by the smoothed level density, 
$\overline{\sum_{ik}\delta(\omega-(\phi_i-\phi_k))}^{\Delta\omega}$. 

\begin{figure}[b]
  \centering
  \includegraphics[width=.6\textwidth]{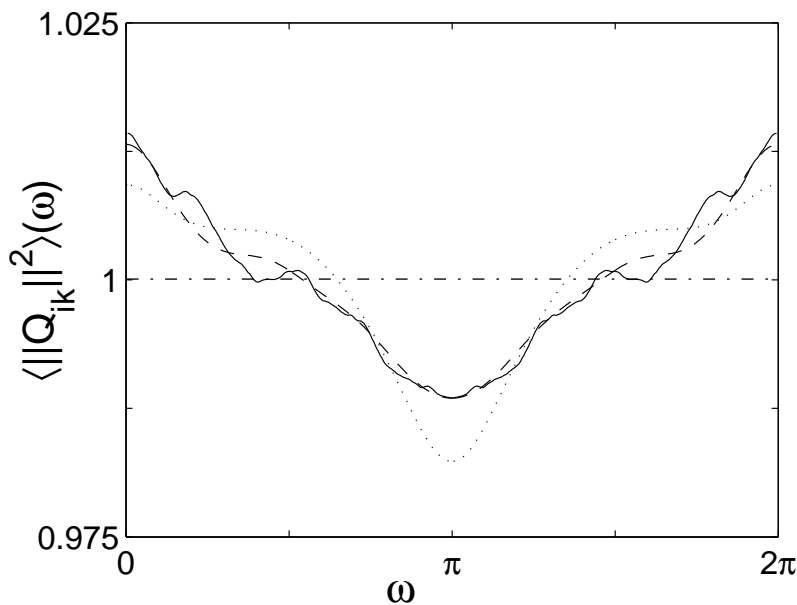}
  \caption{Comparison of averaged overlaps (solid) with semiclassical
  predictions for the baker map on the sphere (full coupling):
  approximate resonances (dashed) and traces (dotted).  Although the
  deviation from RMT (dashed-dotted) 
  is small, semiclassical predictions coincide with
  the quantum result.  Only at $\omega=\pi$ the prediction calculated
  from traces differs slightly.  A single Lorentzian peak cannot be
  observed. Since all resonances have moduli smaller than $\frac12$, the
  corresponding Lorentzians are widely distributed over the interval.}
  \label{fig:hspek0}
\end{figure}
\begin{figure}[p]
  \centering
  \includegraphics[width=.6\textwidth]{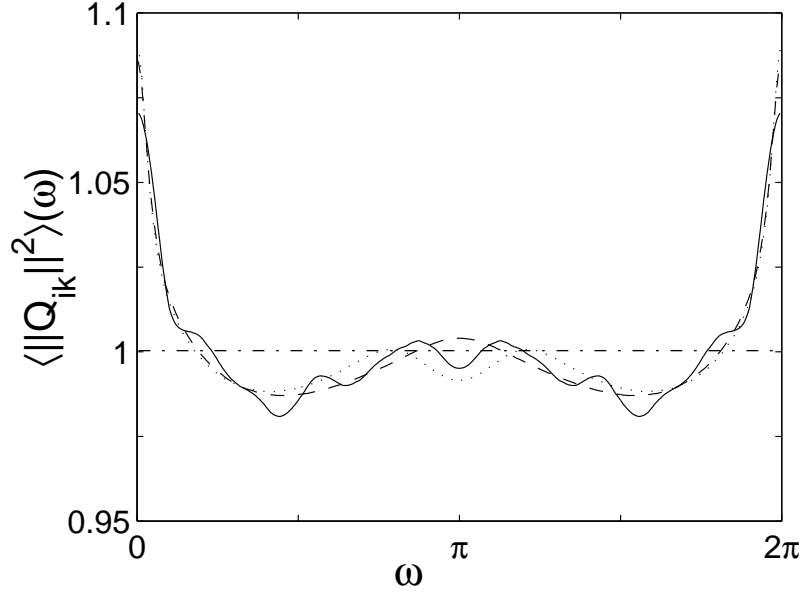}
  \caption{The averaged overlaps and semiclassical predictions for the
  positive coupling (cf.~Fig.~\ref{fig:hspek0}).  At $\omega=0$ we see
  a Lorentzian peak associated to the resonance resulting from small
  coupling. The peak heights for the three cases are nearly equal.}
  \label{fig:hspekp}
\end{figure}
\begin{figure}[p]
  \centering
  \includegraphics[width=.6\textwidth]{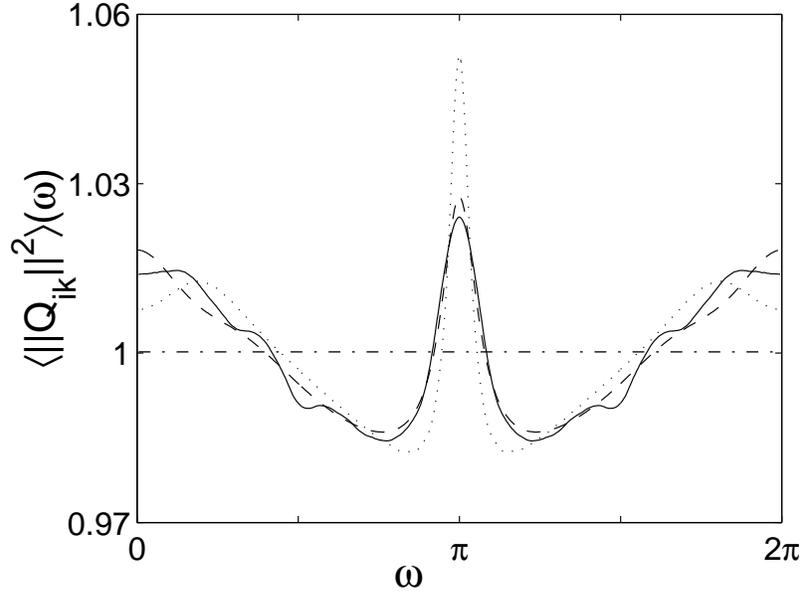}
  \caption{The averaged overlaps and semiclassical predictions for the
  negative coupling (cf.~Fig.~\ref{fig:hspek0}).  The quantum result
  and the semiclassical prediction from approximate resonances coincide,
  while the peak corresponding to the coupling resonance is much sharper
  for the prediction from traces calculated with the use of periodic
  orbits.  This is due to the fact that the approximate resonance in
  question has not reached the final resonance position which is
  calculated analytically (cf.~Sec.~\ref{sec:fixpoints}). Quantum
  uncertainty and noise seem to have the same coarse-graining effect.}
  \label{fig:hspekn}
\end{figure}
As in the classical case Sec.~\ref{sec:traces} we here consider three systems: 
baker map on the sphere and its modifications with positive and negative
coupling.  For all cases we compare our quantum results with both
classical 
predictions, calculated from approximate resonances and from the traces
of the Frobenius--Perron operator obtained with the use of 
periodic orbits, see Fig.~\ref{fig:hspek0}--\ref{fig:hspekn}. 
The averaged overlaps are scaled such that the RMT average 
is equal to unity. 
Up to small fluctuations 
most of the quantum results coincide with the 
semiclassical predictions. 
However, not for every system the quality of the agreement between the semiclassical 
prediction and the quantum results is the same. 
Whereas the agreement is fine for the system with positive coupling, 
the mean resonant peak of the Husimi overlaps approximated by periodic orbits for system (\ref{eq:BSdef}) 
is higher than the peak observed in the quantum results, see Fig.~\ref{fig:hspekn}. 
On the other hand, the prediction derived by classical resonances
obtained by the weak noise approach 
approximate well the quantum results for all systems studied. 
In other words, quantum uncertainty acts similarly as a stochastic perturbation of the classical system. 

\section{Coupled Random Matrices -- Simplification of the Model}
\label{sec:crm}

As it has been explained in the foregoing sections the exchange of 
probability between the two hemispheres is responsible for a resonance. 
We here study a simplification of the system 
of coupled baker maps. The internal dynamics, i.e. the baker maps itself, and their 
corresponding resonances are not the point of interest here. 
Therefore we replace the quantized baker maps (i.e. the
operators~(\ref{eq:hatBN}) and (\ref{eq:hatBS})) by random matrices. 
This simplification might be motivated as follows. 
Consider a strongly chaotic classical system for which 
all classical correlations become arbitrary small already after one 
iteration of the map.  Quantizing such a~system we expect a~random-matrix
like behaviour, 
since all resonances should be located close to the origin and therefore no semiclassical 
corrections will appear. 
A realization of such a system can be constructed by a sequence of several uncoupled 
baker maps $\tilde{B}^k$ (Lyapunov exponent $\approx \ln 2^k$) followed by a coupling 
operator. 
In contrast to the coupled baker maps the coupled random matrices (CRM) 
are more advantageous. 
The resonances are easy to calculate as it will be shown in the sequel. 
Moreover, we have the opportunity to average over an ensemble of
arbitrary many coupled random matrices.

\begin{figure}[t]
  \centering
  \includegraphics[width=0.8\textwidth]{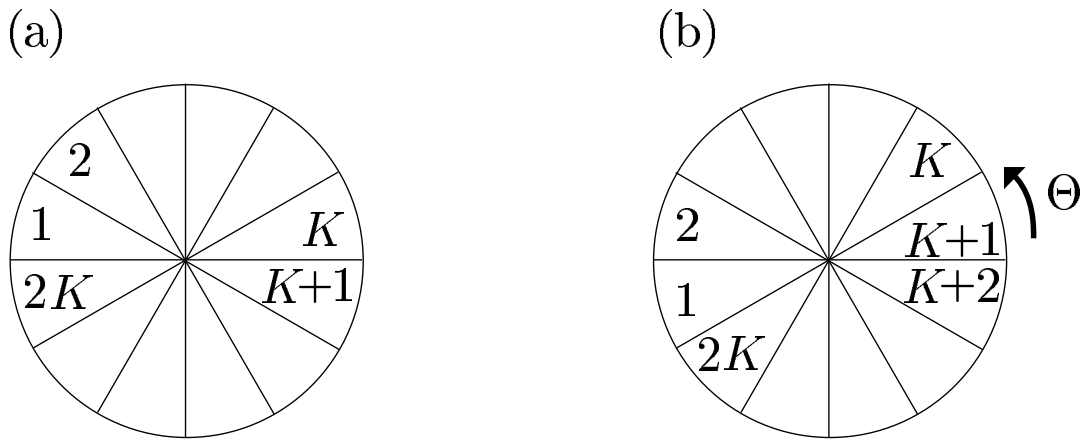}
  \caption{(a) The sphere is partitioned into $2K$ sectors. 
  (b) A rotation of the sphere permutes the sectors. Here the angle 
  $\Theta$ coincides with one sector for simplicity.}
  \label{sector}
\end{figure}

For the classical counterpart of CRM we separate the phase space into two 
partitions of equal area. 
On the sphere we may choose the northern and southern hemisphere. 
A strongly chaotic dynamics acts separately on each hemisphere. The
classical system can be described 
by stochastic matrices. To that end we separate the phase space into $2K$ cells, 
where we conveniently choose $K$ sectors of same area on each hemisphere, 
for instance $1\dots K$ for the northern and $K+1\dots 2K$ for the southern 
hemisphere, see Fig.~\ref{sector}a.
For the $l$--th cell we define a characteristic 
(normalized) density function $\mb l\rb$
which is constant inside the cell and vanishes outside. 
The matrix $A^{(2K)}$ is designed to mimic the FP operator $P$ in this basis  
\begin{equation}
A_{il}^{(2K)}=\lb i\mb P \mb l\rb
\,.
\end{equation}
This is indeed some kind of coarse graining of the Frobenius--Perron operator 
analogous to the Ulam method for the classical system.  
For a strongly chaotic dynamics the matrix elements 
fluctuate around $\frac{1}{K}$, but in this simplification these fluctuations will be 
neglected. Thus the matrix of the uncoupled system $A_0^{(2K)}$ becomes 
\begin{equation}
A^{(2K)}_0=\left(\begin{array}{cc}
w^{(K)}&0\\0&w^{(K)}
\end{array}\right)
\,,
\end{equation}
where $w_{ik}^{(K)}=\frac{1}{K}$. 
The eigenvalues of this matrix are easily calculated as $(1, 1, 0,\dots, 0)$, 
where the degeneracy of eigenvalue $1$ is due to the disjoint 
invariant densities on each hemisphere. 

To generate a non-zero resonance we introduce a coupling between both 
hemispheres as a rotation of the sphere by $\Theta$ around the $x$ axis
(perpendicular to the plane of Fig.~\ref{sector}).
The system size is chosen that $\frac{\Theta}{2\pi}2K=k$ is integer. 
In this representation the rotation becomes a $k$-fold cyclic permutation 
of the cells. 
Then the (coupled) system is described by 
\begin{equation}
A_{\Theta}^{(2K)}=C_{\Theta}^{(2K)}A_0^{(2K)}
\,,
\end{equation}
where $C_{\Theta}^{(2K)}$ is a permutation matrix acting as 
$i\rightarrow i=i+k \bmod 2K$, see Fig.~\ref{sector}b. 
A $k$-fold permutation of cells acts on the block-diagonal matrix 
like a $k$-fold shift of the row vectors, where the trace of this matrix 
changes for each permutation by $-\frac{2}{K}$ as long as $k$ is smaller than $K$. 
The eigenvalues of this shifted matrix are easily calculated. 
Since the matrix is still of rank 2, $2K-2$ eigenvalues are equal to
zero.
One eigenvalue must be 1, because the matrix is double stochastic. 
Thus the missing eigenvalue can be calculated from the trace as 
$\lambda_1={\rm tr} A^{(2K)}-1$. 
Finally the eigenvalue becomes $\lambda_1=\frac{2K-2k}{K}-1=1-\frac{2\Theta}{\pi}$. 

By varying the rotation angle $\Theta$ the second resonance is controllable over 
the range $[-1,1]$. For $\Theta=0$ the eigenvalue $1$ is degenerated, since 
we have two uncoupled systems. 
A rotation with $\Theta=\frac{\pi}{2}$ mixes both sides uniformly, such that 
the resonance goes to 0. 
Finally, choosing $\Theta=\pi$ the hemispheres exchange completely
after one map, whereby the resonance 
becomes $-1$.

The quantization of such a system is easily done. 
The separated chaotic dynamics on each hemisphere correspond to a 
Floquet matrix which is block diagonal in the usual $|j,m\rangle$ basis. 
Due to the phase-space partition we use even dimensions of the Hilbert space, 
otherwise we are not able to 
assign the state $|j,m=0\rangle$ to one of the hemispheres. 
The block matrices are chosen as random 
unitary matrices due to the strongly chaotic 
dynamics. The coupling becomes a rotation matrix
$R_x(\Theta)=e^{-i\Theta\hat J_x}$ 
\begin{equation}
U=R_x(\Theta)\left(\begin{array}{cc}
U_1&0\\0&U_2\end{array}\right) 
\,,
\end{equation}
and $U_{1,2}$ are independent random unitary matrices distributed
according to the Haar measure on $U(N/2)$.

\begin{figure}[p]
  \centering
  \includegraphics[width=.6\textwidth]{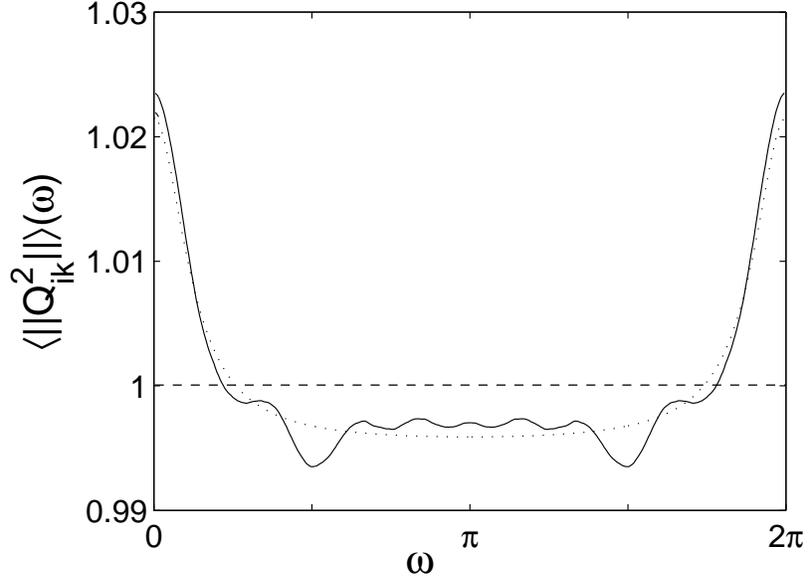}
  \caption{Comparison of averaged overlaps 
  with the semiclassical prediction for CRM with positive
  coupling.  For the quantum results (solid) we find 
  good agreement with the semiclassical prediction (dotted).
  Uniform spectrum predicted by RMT is denoted by a~dashed line.}
  \label{fig:rmtpos}
\end{figure}
\begin{figure}[p]
  \centering
  \includegraphics[width=.6\textwidth]{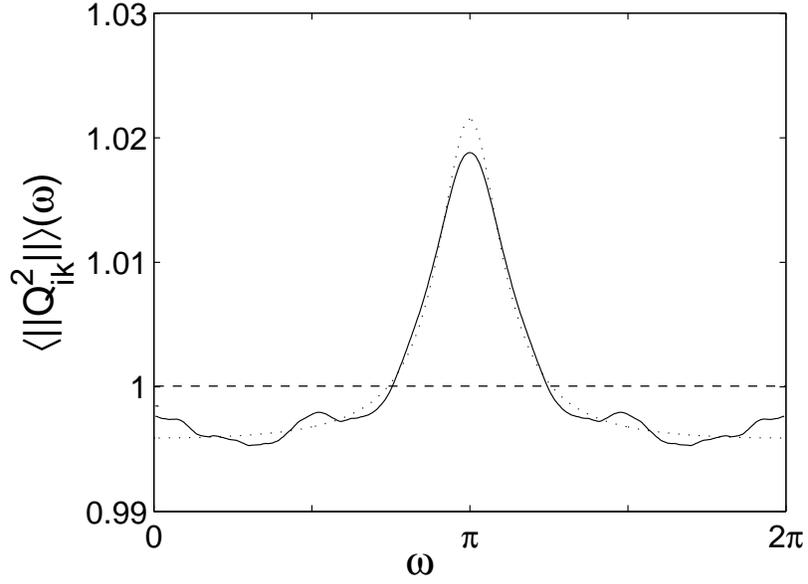}
  \caption{Averaged overlaps of CRM and semiclassical prediction 
  for negative coupling (cf.~Fig.~\ref{fig:rmtpos}).}
  \label{fig:rmtneg}
\end{figure}
We choose two different coupling angles $\Theta=0.5$ and $\Theta=\pi+0.5$, 
where the resonances become 
$\lambda_\pm=\pm(1-\frac{1}{\pi})$, respectively.  
For each case the Hilbert space dimension is $N=200$. 
Due to the semiclassical prediction the averaged overlaps distribute as
\begin{equation}
\left\langle||Q_{ik}||^2\right\rangle(\omega)\propto
1+\frac{1}{N-1}
\sum_{n=1}^{\infty}\lambda_{\pm}^n\cos(n\omega)
\,.
\end{equation}
As in the case of the coupled baker maps we use the smoothing with a sinc function of 
form $\sin(10\omega)/\omega$. Fig.~\ref{fig:rmtpos} shows the 
averaged overlaps, 
$\left\langle||Q_{ik}||^2\right\rangle(\omega)$, 
for the positive resonance $\lambda_+$, while 
in Fig.~\ref{fig:rmtneg}
we see the results for the negative resonance $\lambda_-$. 
For both cases we find a good agreement with the semiclassical prediction. 
The small fluctuations (much smaller than the Lorentz peaks) of the
quantum results are non-generic, i.e. they look different if one chooses
other pairs of random matrices. 

\section*{Concluding remarks}

We investigate classical systems whose Frobenius-Perron 
operators have  resonances with large moduli. 
Such long-lived excitations have a strong impact 
on the corresponding quantum dynamics. 
In particular, they constitute a scarring mechanism for 
quantum eigenfunctions. 

In order to detect resonances we employed the small-additive-noise approach. 
The noise was designed such as to ensure a finite-dimensional 
representation of the Frobenius-Perron operator which has a 
unique discrete spectrum. 

Our models are chosen such that the classical dynamics is 
well understood --- their topological and metric entropies are both
equal to $\ln2$.
For all maps analyzed we provide Markov partitions, 
as well as the complete sets of periodic orbits. 
Making use of them we could evaluate the classical 
Cvitanovi\`c-Eckhardt  
trace formula  
for the Frobenius-Perron operator \cite{cvit2} 
and discuss its relation to 
the quantum return probabilities  with respect to coherent states. 
That return probability naturally leads to an intimate 
relation between long-lived classical resonances 
and quantum scars. 
In fact, system specific quantum localization properties 
not explained by the standard ensembles of random matrices 
turn out interpretable statistically on the basis of 
classical Ruelle-Pollicott resonances.

\begin{acknowledgments}
We would like to thank to Prot Pako\'nski for many fruitful
discussions.  Financial support by the Polish State Committee for
Scientific Research (KBN) Grant No 5~P03B~018~21 and the
Sonderforschungsbereich 237 der Deutschen Forschungsgemeinschaft is
gratefully acknowledged.
\end{acknowledgments}

\appendix
\section{Exemplary transition matrices}\label{app:Tmat}
We present here the transition matrices defined in
Sec.~\ref{sec:fixpoints} for system (\ref{eq:BNdef}) and
(\ref{eq:BSdef}) for coupling parameter $\Delta=1/8$.  Assuming the
lexicographical ordering of cells (that is $A_1,A_2,B_1,B_2,\ldots$) the
transition matrix $\mathbf{T^+}$ for the system~(\ref{eq:BNdef}) with
positive coupling is equal
\begin{equation}\label{poscoup}
  \mathbf{T^+} = \frac18\left(
  \begin{array}{*{8}{c}}
  0 & 4 & 0 & 4 & 0 & 0 & 0 & 0 \\
  4 & 0 & 4 & 0 & 0 & 0 & 0 & 0 \\
  0 & 0 & 0 & 0 & 8 & 0 & 0 & 0 \\
  0 & 0 & 0 & 0 & 0 & 8 & 0 & 0 \\
  0 & 0 & 0 & 0 & 0 & 0 & 8 & 0 \\
  0 & 0 & 0 & 0 & 0 & 0 & 0 & 8 \\
  1 & 0 & 1 & 0 & 2 & 0 & 4 & 0 \\
  0 & 1 & 0 & 1 & 0 & 2 & 0 & 4
  \end{array}
  \right)\ ,
\end{equation}
while for the negative coupling~(\ref{eq:BSdef}) it reads
\begin{equation}\label{negcoup}
  \mathbf{T^-} = \frac18\left(
  \begin{array}{*{8}{c}}
  0 & 4 & 0 & 2 & 0 & 1 & 0 & 1  \\
  4 & 0 & 2 & 0 & 1 & 0 & 1 & 0  \\
  0 & 8 & 0 & 0 & 0 & 0 & 0 & 0  \\
  8 & 0 & 0 & 0 & 0 & 0 & 0 & 0  \\
  0 & 0 & 0 & 8 & 0 & 0 & 0 & 0  \\
  0 & 0 & 8 & 0 & 0 & 0 & 0 & 0  \\
  0 & 0 & 0 & 0 & 4 & 0 & 4 & 0  \\
  0 & 0 & 0 & 0 & 0 & 4 & 0 & 4 
  \end{array}
  \right)\ .
\end{equation}
It is straightforward to calculate their eigenvalues and in particular
one gets that the second largest eigenvalue in the case of positive
coupling (\ref{poscoup})
is equal $\lambda_2^+=\frac{a}6+\frac1a\approx0.8846$ where
$a=\sqrt[3]{27+3\sqrt{57}}$, while in the case of negative coupling
(\ref{negcoup}) one has 
$\lambda_2^- = -\lambda_2^+$.

\section{Contribution of poles to the traces of FP operator}\label{app:poles}

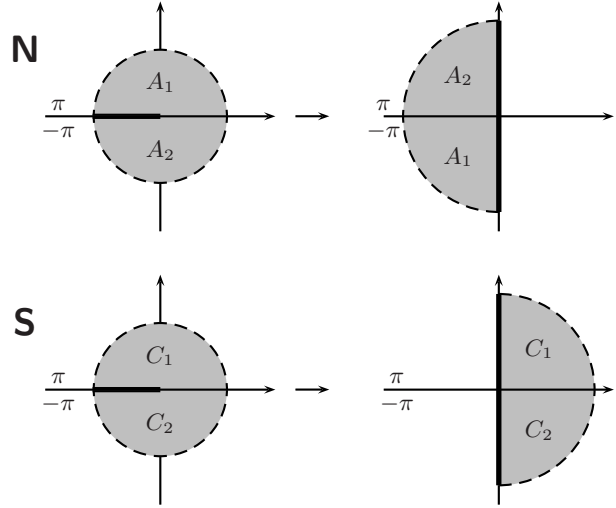
\begin{figure}
  \centering
  \psset{unit=.9cm}
  \vbox{
  \begin{pspicture}(0,0)(9,4)
    \psset{fillstyle=solid,fillcolor=lightgray}
    \rput(0,3){\Large\bfseries\sffamily N}
    \rput(2,2){
      \psline{->}(0,-1.7)(0,1.7)
      \pscircle[linestyle=dashed](0,0){1} \psline[linewidth=2pt](-1,0)(0,0)
      \psline{->}(-1.7,0)(1.7,0)
      \rput[b](-1.5,.1){$\pi$} \rput[t](-1.5,-.1){$-\pi$}
      \rput(0,.5){$A_1$} \rput(0,-.5){$A_2$}
    }
    \psline{->}(4,2)(4.5,2)
    \rput(7,2){
      \psline{->}(0,-1.7)(0,1.7)
      \psarc[linestyle=dashed](0,0){1.4142}{90}{270}
      \psline{->}(-1.7,0)(1.7,0)
      \psline[linewidth=2pt](0,-1.4142)(0,1.4142)
      \rput[b](-1.7,.1){$\pi$} \rput[t](-1.7,-.1){$-\pi$}
      \rput(-.6,.6){$A_2$} \rput(-.6,-.6){$A_1$}
    }
  \end{pspicture}\\
  \begin{pspicture}(0,0)(9,4)
    \psset{fillstyle=solid,fillcolor=lightgray}
    \rput(0,3){\Large\bfseries\sffamily S}
    \rput(2,2){
      \psline{->}(0,-1.7)(0,1.7)
      \pscircle[linestyle=dashed](0,0){1} \psline[linewidth=2pt](-1,0)(0,0)
      \psline{->}(-1.7,0)(1.7,0)
      \rput[b](-1.5,.1){$\pi$} \rput[t](-1.5,-.1){$-\pi$}
      \rput(0,.5){$C_1$} \rput(0,-.5){$C_2$}
    }
    \psline{->}(4,2)(4.5,2)
    \rput(7,2){
      \psline{->}(0,-1.7)(0,1.7)
      \psarc[linestyle=dashed](0,0){1.4142}{-90}{90}
      \psline{->}(-1.7,0)(1.7,0)
      \psline[linewidth=2pt](0,-1.4142)(0,1.4142)
      \rput[b](-1.5,.1){$\pi$} \rput[t](-1.5,-.1){$-\pi$}
      \rput(.6,.6){$C_1$} \rput(.6,-.6){$C_2$}
    }
  \end{pspicture}
  }
  \caption{The evolution of the phase space around the north and south
  poles governed by the maps (\ref{eq:N_pole}) and (\ref{eq:S_pole}) ---
  top and bottom row, respectively.}
  \label{fig:poles}
\end{figure}

In order to evaluate the contribution to the trace
formula~(\ref{eq:trace}) of the poles we first
consider stereographic projections of their neighborhoods.  In case of
the north pole we use stereographic projection from the south pole and
vice versa so that in each case the pole corresponds to the origin of the
plane.  Next we introduce coordinate system such that the discontinuity
line of both points corresponds to the negative $x$ axis.  Using polar
coordinates with $\varphi\in[-\pi,\pi]$ we can easily express the action
of the map (the behaviour around the poles of the ``positive'' and
``negative'' version is the same) and the mapping around the north pole
is given by
\begin{equation}\label{eq:N_pole}
  \left\{\begin{array}{c}
    \varphi \longrightarrow \varphi' = \varphi/2 + \pi\\
    r \longrightarrow r' = \sqrt{2}\,r
  \end{array}\right.\ ,
\end{equation}
while for the south pole it reads
\begin{equation}\label{eq:S_pole}
  \left\{\begin{array}{c}
    \varphi \longrightarrow \varphi' = \varphi/2\\
    r \longrightarrow r' = \sqrt{2}\,r
  \end{array}\right.\ .
\end{equation}
The action of these maps is presented in Fig.~\ref{fig:poles}.
From these equations it is clear that the expression~(\ref{eq:trace}) is
not well defined at the poles.  In order to regularize it we replace the
$\delta$ function with a~two-dimensional Gaussian of width $\sigma$,
compute the integral and later go to the limit $\sigma \rightarrow 0$.
To this end we need square of the distance between the initial point and its
$n$-th iteration
\begin{equation}
   d^2(r,\varphi) = r^2\left[ 1 + 2^n \mp
    2(\sqrt2)^n\cos(1-2^{-n})\varphi \right]
\end{equation}
where the upper and lower sign is for the north and south pole,
respectively.  Putting it all together we get that the contribution to
the trace of $P^n$ coming from the pole is
\begin{equation}
  \lim_{\sigma \rightarrow 0} \frac1{2\pi\sigma^2}
    \int_{-\pi}^\pi {\rm d}\varphi\int_0^\infty {\rm d}r\, r \exp\left(-\frac{
    d^2(r,\varphi)}{2\sigma^2}\right)\ .
\end{equation}
Performing first the radial integral we get
\begin{equation}\label{eq:phiinteg}
  \frac1{2\pi}\int_{-\pi}^\pi {\rm d}\varphi \frac1{1 + 2^n \mp
    2(\sqrt2)^n\cos(1-2^{-n})\varphi}
\end{equation}
which does not depend on the width $\sigma$.  Hence the limit $\sigma
\rightarrow 0$ is automatically obtained.  The
integral~(\ref{eq:phiinteg}) is elementary and one gets the following
result for the contribution of the poles
\begin{equation}
  b_\mp(n) = \frac2\pi \frac{2^n}{(2^n-1)^2} \arctan\left(
    \frac{(\sqrt2)^n \mp 1}{(\sqrt2)^n \pm 1}
    \tan(1-2^{-n})\frac\pi 2
  \right)
\end{equation}
where the upper and lower sign is for the north and the south pole,
respectively.  If one compares the above values with the contributions
of unstable and inverse unstable fixpoints then one finds that the south
pole can be nearly treated like the unstable fixpoint while the north
pole differs a~bit from the inverse unstable one.  To get explicitly the
expression~(\ref{eq:trace}) for the traces one has to count the number of
periodic points using the connectivity matrix $\mathsf{T}$.  For
instance, for map~(\ref{eq:BSdef}) we obtain
\begin{equation}
  \tr P^n = \frac{\tr(\mathsf{T^-})^n-4-2(-1)^n}{2^n+2^{-n}-2} + b_+(n) + b_-(n) \ ,
\end{equation}
where we used~(\ref{eq:pp_trT}) and taken into account that Markov
partition does not feel the topology of the system (the lines
$t=\cos\theta=\pm1$ are single points --- poles) and also the period
two orbit originating from $\varphi=\pi$ and $t=-1/3$ is counted twice
(see Sec.~\ref{sec:fixpoints} and also left hand side of the
Fig.~\ref{fig:husimi}).  Analogously for map~(\ref{eq:BNdef}) we have
\begin{equation}
  \tr P^n = \frac{\tr(\mathsf{T^+})^n-3-(-1)^n}{2^n+2^{-n}-2} + b_+(n) + b_-(n) \ .
\end{equation}
In case of the standard baker map on the sphere ($\Delta=1/2$ case of
maps~(\ref{eq:BNdef}) and (\ref{eq:BSdef})) we are able to find the
analytical expression for the number of periodic points so the traces of
FP operator are given by
\begin{equation}
  \tr P^n = \frac{2^n-3-(-1)^n}{2^n+2^{-n}-2} + b_+(n) + b_-(n) \ .
\end{equation}

\end{document}